\documentclass[journal,twocolumn]{IEEEtran}
\usepackage{amsmath,amssymb,amsfonts}
\usepackage[noadjust]{cite}
\usepackage{graphicx}
\usepackage{epstopdf}
\usepackage[noend]{algpseudocode}
\usepackage{algorithmicx,algorithm}
\usepackage{float}
\usepackage{array}
\usepackage{color}
\usepackage{verbatim}
\usepackage{subeqnarray}
\usepackage{cases}
\usepackage{lettrine}
\usepackage{sidecap}
\usepackage{enumitem}
\usepackage{colortbl}
\usepackage{lineno}
\usepackage{extpfeil}

\ifCLASSOPTIONcompsoc
\usepackage[caption=false,font=normalsize,labelfont=sf,textfont=sf]{subfig}
\else
\usepackage[caption=false,font=footnotesize]{subfig}
\fi

\makeatletter
\newcommand{\removelatexerror}{\let\@latex@error\@gobble}
\makeatother
\def\BibTeX{{\rm B\kern-.05em{\sc i\kern-.025em b}\kern-.08em		
		T\kern-.1667em\lower.7ex\hbox{E}\kern-.125emX}}
\begin{document}\bibliographystyle{IEEEtran}
\title{Coexistence Designs of Radar and Communication Systems in a Multi-path Scenario}
\author{Haoyu Zhang,~Li Chen,~\IEEEmembership{Senior Member,~IEEE}, Kaifeng Han,~Yunfei Chen,~\IEEEmembership{Senior Member,~IEEE}, and Guo Wei
\thanks{Haoyu Zhang, Li Chen and Guo Wei are with the CAS Key Laboratory of Wireless-Optical Communications, University of Science and Technology of China, Hefei 230027, China (e-mail: hyzhangy@mail.ustc.edu.cn; \big\{chenli87, wei\big\}@ustc.edu.cn).}

\thanks{Kaifeng Han is with the China Academy of Information and Communications Technology, Beijing 100191, China (e-mail: hankaifeng@caict.ac.cn).}

\thanks{Yunfei Chen is with the School of Engineering, University of Warwick, Coventry CV4 7AL, U.K. (e-mail: yunfei.chen@warwick.ac.uk).}
}
\maketitle
\begin{abstract}
The focus of this study is on the spectrum sharing between multiple-input multiple-output (MIMO) communications and co-located MIMO radar systems in multi-path environments. The major challenge is to suppress the mutual interference between the two systems while combining the useful multi-path components received at each system. We tackle this challenge by jointly designing the communication precoder, radar transmit waveform and receive filter. Specifically, the signal-to-interference-plus-noise ratio (SINR) at the radar receiver is maximized subject to constraints on the radar waveform, communication rate and transmit power. The multi-path propagation complicates the expressions of the radar SINR and communication rate, leading to a non-convex problem. To solve it, a sub-optimal algorithm based on the alternating maximization is used to optimize the precoder, radar transmit waveform and receive filter iteratively. Simulation results are provided to demonstrate the effectiveness of the proposed design.
\end{abstract}
\section{Introduction}
\lettrine[lines=2]{T}{HE} explosive growth of mobile devices has placed an urgent demand for exploring extra radio spectrum resources. To cope with that, the frequency bands traditionally occupied by radar systems are shared with the wireless communication systems, e.g., sub-6 GHz band leveraged by air traffic control radars and long-range weather radars \cite{6967722}, and the millimeter wave (mmWave) band  conventionally assigned to automotive radars and high-resolution imaging radars \cite{7786130}. This has led to strong interest in the coexistence of radar and communication (CRC) \cite{9393464}.

The primary challenge of CRC is to manage the interference at both radar and communication subsystems. Early studies included policies for opportunistic spectrum access \cite{6331681}, separate radar waveform \cite{6933960,6558027} and communication receiver \cite{8332962} design. The joint design of the radar and communication signals increases the degrees of cooperation between the two systems, thus suppressing the interference more effectively. The cooperative spectrum sharing between a single-input single-output (SISO) pulsed radar and a SISO communication system was considered in \cite{8048004,9359480}. Specifically, in \cite{8048004}, the performance metric of the communication system was formulated as a weighted sum of rates with and without the radar interference, which is then maximized with constraints on radar signal-to-interference-plus-noise-ratio (SINR) and transmit power. The work of \cite{9359480} investigated radar-oriented multi-carrier CRC systems and developed optimum power allocation strategies to maximize the radar SINR subject to communication throughput and power constraints.

For the CRC systems based on multiple-input multiple-output (MIMO) architectures,
cooperative mutual interference management methods in the space-time domain were proposed in \cite{7470514,8477186}. In \cite{7470514}, the communication codebook was jointly designed with MIMO-matrix completion (MC) radar sampling scheme to minimize the effective interference power of the radar system, while the average capacity and transmit power constraints on the communication system were taken into account. The authors in \cite{8477186} borrowed the interference alignment methods from communication theory and jointly designed the transmit and receive beamformers for both the MIMO radar and MIMO communication systems to eliminate mutual interference. A more practical multiple-antenna scenario was considered in \cite{8352726,9582836}, where the radar waveform was chosen to bear the requirements concerning range resolution, sidelobe level and envelope constancy. In \cite{8352726}, the radar SINR at a single resolution cell was maximized subject to the constraints on the transmit power, the communication rate and the similarity of the radar waveform to a reference signal. In \cite{9582836}, the figure of merit was constructed as the inverse of the harmonic mean of the radar SINR across multiple resolution cells, which was minimized under the constraints similar to those in \cite{8352726}.

{The aforementioned contributions generally assume that the CRC systems detect the target via a direct line-of-sight (LoS) path and serve communication users in rich scattering environments. Recently, to meet the demand for more spectral and spatial resources, both radar and communication systems have evolved towards the same direction of large bandwidth and massive antennas, which makes physical channels exhibit a sparse multi-path structure \cite{bajwa2009sparse,masood2014efficient}. Although dispersed signals from Non-LoS (NLoS) paths may damage communication and radar performance, they also provide more degrees of freedom (DoFs). Specifically, for communication, since NLoS propagation provides diversity and multiplexing gains, it is necessary to combine multi-path components to enhance data transmission performance \cite{7343329,7504275}. For radar, the useful target multi-path returns include not only the backscattered signal along LoS, but also echoes from reflectors. By exploiting the geometric relationship between the NLoS paths and the reflectors in the resolvable multi-path components, the radar system can combine the multi-path channels to change them from invalid interference to effective signals, thus enhancing the detection capability \cite{7001702,zhang2021moving,9562316}.}

{All these techniques are developed for multi-path combining of individual radar or communication systems. However, for the CRC systems in sparse multi-path environments, it becomes more complex. Specifically, the multi-path combining of each subsystem is for drastically different purposes, and the existence of mutual interference poses greater challenges on it. As a consequence, the known multi-path combining designs for radar-only or communication-only are inapplicable for the coexistence scenario.} To the best of our knowledge, combining useful multi-path signals from individual systems and simultaneously suppressing the mutual interference between the two systems for better detection performance and higher data rates has not been investigated before.

Motivated by the above observations, in this work, we consider the coexistence between a co-located pulsed MIMO radar and a single-user MIMO communication system in sparse multi-path environments. The communication precoder, radar transmit waveform and receive filter are jointly designed to achieve interference suppression and multi-path combining. Specifically, we take the SINR and the transmission rate as the performance metrics of the radar system and the communication system, respectively. In order to effectively utilize the diversity gain brought by multi-path propagation, the multi-path returns of the target are exploited to further improve the SINR. The existence of the mutual interference makes the simultaneous optimization of the SINR and rate difficult. To mitigate the impact of interference on the considered CRC system, we formulate the radar-centric design strategy, where the radar SINR is maximized with constraint on the communication transmission rate. The resulting problem is complex due to the non-convex constraints and non-concave objective function. We adopt the alternating optimization method to decouple the optimization variables and propose effective algorithms to solve each non-convex sub-problems. The main contributions are summarized as follows.

\begin{itemize}[topsep = 0 pt]	
	\item \textbf{Multi-path CRC model:} {Different from previous works \cite{7470514,8477186,8352726,9582836}, we consider the MIMO CRC systems in sparse multi-path environments and construct the corresponding multi-path signal model. Each subsystem suffers from not only the interference form its counterpart but also the multi-path propagation of its own transmission. We generalize the traditional performance metrics of radar SINR and communication transmission rate to multi-path situations. The suppression of mutual interference and multi-path combining of useful signals are necessary to enhance the performance of each subsystem.}

	\item \textbf{Communication precoder design:} Compared with the communication design in \cite{7470514,8477186,8352726,9582836}, our constructed sub-problem on the communication precoder is more intractable due to multi-path propagation and dynamic radar interference. We exploit the {successive convex approximation} (SCA) method to 	approximate the sub-problem as a series of convex {quadratically constrained quadratic programming} (QCQP) problems. Further, the {alternating direction method of multipliers} (ADMM) method is utilized to accelerate the solution of each QCQP problem in the SCA procedure.

	\item \textbf{Radar waveform design:} In contrast to the waveform design for LoS detection, we consider the exploitation of multi-path target echoes and incorporate common waveform constraints into the sub-problem. Specifically, we use the SCA method to approximate the sub-problem as a series of fractional {semi-definite programming} (SDP) problems. When considering the similarity constraint, we construct an equivalent problem of each fractional SDP problem to solve it optimally. When considering the peak-to-average power ratio (PAPR) constraint, we use the alternate projection to efficiently obtain the solution of each fractional SDP problem.

\end{itemize}

The remainder of the paper is organized as follows. Section II introduces the system model and formulates the optimization problem. The SCA-based algorithm for obtaining the communication precoder is presented in Section III. Section IV develops effective radar waveform design algorithms under similarity and PAPR constraints, respectively. The numerical results are given in Section V, followed by the conclusions drawn in Section VI.

\emph{Notations: }We use bold lowercase letters to represent column vectors, and bold uppercase letters to represent matrices. The operators ${{\left( \cdot  \right)}^{T}}$, ${{\left( \cdot  \right)}^{*}}$ and ${{\left( \cdot  \right)}^{H}}$ correspond to the transpose, conjugate and Hermitian transpose, respectively. ${\rm tr}\left( \textbf{A}  \right)$, $\left| \textbf{A} \right|$, ${\rm rank}\left( \textbf{A}  \right)$ and ${\rm vec}(\textbf{A})$ stand for the trace, the determinant, the rank and the vectorization operation of the matrix $\textbf{A}$, respectively. $\mathcal{M}\left( \textbf{A} \right)$ denotes the normalized principle eigenvector of $\textbf{A}$. $\textbf{A}\succ \textbf{0}$ ($\textbf{A}\succcurlyeq \textbf{0}$) indicates that $\textbf{A}$ is positive definite (positive semidefinite). $\left\| \textbf{a} \right\|$ is the Euclidean norm of the vector $\textbf{a}$. ${\textbf{I}_{M}}$ is a $M \times M$ identity matrix. The symbol $\otimes$ represents the Kronecker product. ${\mathbb{C}}^{M \times N}$ denotes the set of complex-valued $M \times N$ matrices. $\Re \left\{ x \right\}$ refers to the real parts of a complex number $\emph{x}$. $\mathbb{E}(\cdot)$ denotes the statistical expectation. Finally,  $\textbf{x}\sim \mathcal{CN}\left( \textbf{a},{\textbf{A}} \right)$ means that $\textbf{x}$ follows a complex Gaussian distribution with mean $\textbf{a}$ and covariance matrix $\textbf{A}$.

\section{System Model and Problem Formulation}

\begin{figure}[!t]
	\centering
	\includegraphics[width=0.5\textwidth]{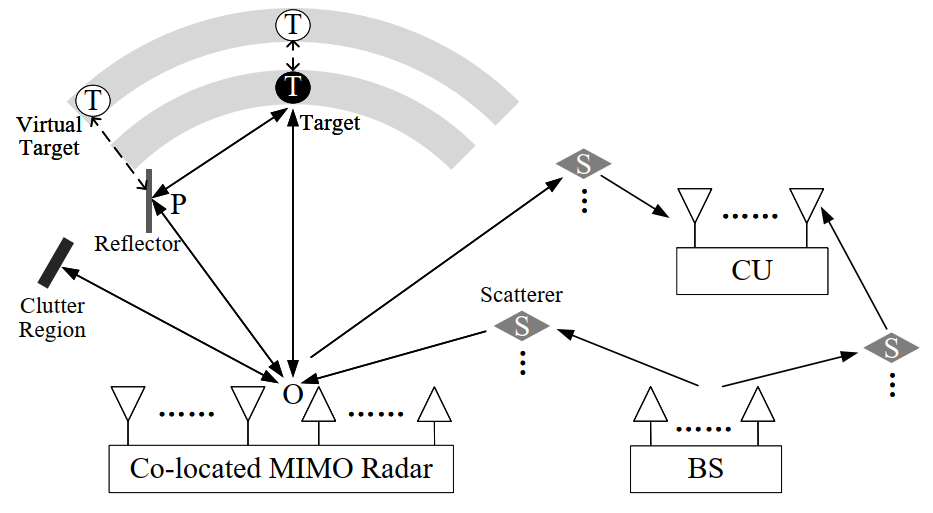}	
	\caption{{Spectrum sharing between a co-located MIMO radar and a MIMO communication system in multi-path environments.}}\label{system_model}
\end{figure}

We consider the scenario illustrated in Fig. \ref{system_model}, where a co-located pulsed MIMO radar operates on the same frequency band as a single-user MIMO communication system. The communication system is composed of a {base station} (BS) and a {communication user} (CU) equipped with $N_T$ and $N_R$ antennas, respectively, where the BS serves the CU with desired rates in the presence of multi-path propagation generated by far-field scatterers. The radar system is equipped with $M_T$ transmit and $M_R$ receive antennas, {which detects a point-like target in the presence of diffuse multi-path.} All antennas are assumed to be deployed in half-wavelength spaced uniform linear arrays. 

The radar pulse signal is transmitted at a duration of $K/B$, where $B$ denotes the sampling rate of the radar system. A burst of $P$ pulses are transmitted at a {pulse repetition interval} (PRI) of $\tilde{K}/B$ in a {coherent processing interval} (CPI), and $P$ is chosen in such a way that no cell migration takes place in the CPI \cite{8048004}. Denote $\textbf{s}\left( k \right)\in {{\mathbb{C}}^{{{M}_{T}\times 1}}}$ as the discrete-time transmitted signal at time instant $k$ for $\big( k\text{ mod }\tilde{K} \big)\in \left\{ 1,\ldots ,K \right\}$. Then, the transmit radar waveform during each PRI can be represented by $\textbf{S}=\left[ \textbf{s}\left( 1 \right)\cdots \textbf{s}\left( K \right) \right]\in {{\mathbb{C}}^{{{M}_{T}}\times K}}$.

For the communication system, the BS continuously transmits communication symbols to the CU during the CPI\footnote{{The clocks at the BS and the radar are updated periodically, so that the clock offset between them can be ignored. The clock information of the CU can be fed back to the radar through the BS. The radar receiver and the CU sample their respective received signals with the same sampling rate at the agreed time, thus realizing the synchronization of the sampling time \cite{8477186}.}}. The symbol vector transmitted at time instant $k$ is denoted by $\textbf{d}\left( k \right)\in {{\mathbb{C}}^{D\times 1}}$ with $D$ being the number of data streams and satisfying $D\le \min \left\{ {{N}_{T}},{{N}_{R}} \right\}$. These symbols are assumed to be independent and identically Gaussian distributed, i.e., $\textbf{d}\left( k \right)\sim \mathcal{C}\mathcal{N}\left( \textbf{0},{\textbf{I}_{D}} \right),\forall k$ {\cite{tse2005fundamentals}}. Denote the precoder for the BS as $\textbf{V}\in {{\mathbb{C}}^{{{N}_{T}}\times D}}$. After precoder, the symbol vectors for the CU become $\textbf{x}\left( k \right)=\textbf{V}\textbf{d}\left( k \right)\in {{\mathbb{C}}^{{{N}_{T}\times 1}}},\forall k$.

\subsection{Radar Performance}
As shown in Fig. \ref{system_model}, {the signal backscattered from the target is received at the radar via a direct LoS path and several indirect NLoS paths. In general, multi-path propagation may create virtual targets, which needs to be distinguished from the real target using their different features such as distance and signal strength, etc \cite{feng2021multipath,dong2021through}. If the geometric characteristic of the reflector are known, the multi-path propagation can be predicted and the target echoes along NLoS paths can be exploited to enhance the radar sensing. Here we mainly consider the exploitation of first-order reflection that can be completely separated from the direct target echo in the fast-time domain. In addition, the radar is also subject to clutter and communication interference from different regions. To simplify the exposition and without loss of generality, we analyze the radar operation in the $p$-th PRI.

Assume that the radar pulse hits the target located in the angular direction of ${{\theta }_{0}}$ with a round-trip delay $2{\tau }_{0}/B$ along the LoS, i.e., the target will be present in an unknown range cell ${\tau }_{0}$ with ${\tau }_{0} \in \big\{ {K},\ldots ,{ \tilde{K}-K } \big\}$. With aid of the reflection surface, the radar can also detect the target along the NLoS. {The scattering from the rough surface can be modeled by multiple spatially distributed reflections from the patches that make up the surface. Each reflection is associated with the formation of a bistatic configuration of "radar"-"scattering patch"-"target", resulting in two types of indirect path return (see $\overrightarrow{\rm OP}{\rightarrow}\overrightarrow{\rm PT}{\rightarrow}\overrightarrow{\rm TO}$ and $\overrightarrow{\rm OT}{\rightarrow}\overrightarrow{\rm TP}{\rightarrow}\overrightarrow{\rm PO}$ in Fig. \ref{system_model}). Moreover, signals from different patches are statistically independent from each other \cite{7001702}.} Suppose that there exists $J$ patches causing the target returns along NLoS paths, where the $j$-th patch has an angle of ${{\theta }_{j}}$ to the radar receiver and its corresponding NLoS paths have a relative propagation delay of ${{\tau }_{j}}$ compared with the target echo along the LoS. After aligning the time delays of each path, the multi-path returns of the target, denoted by ${\textbf{Y}_{m}}$, can be formulated as\footnote{Possible Doppler phase variations can be ignored within a PRI, since one single PRI duration $\tilde{K}/B$ is generally short \cite{9001239}.}
\begin{equation}
\begin{aligned}
\textbf{Y}_{m}=&{{\alpha }_{0}}{\textbf{b}_{r}}\left( {{\theta }_{0}} \right)\textbf{b}_{t}^{T}\left( {{\theta }_{0}} \right)\textbf{S}+\sum\limits_{j=1}^{J}{    {{{\alpha }_{j}}} \big[ {\textbf{b}_{r}}\left( {{\theta }_{0}} \right)\textbf{b}_{t}^{T}\left( {{\theta }_{j}} \right)}+\\
&{\textbf{b}_{r}}\left( {{\theta }_{j}} \right)\textbf{b}_{t}^{T}\left( {{\theta }_{0}} \right) \big]\textbf{S},\label{radar_channel}
\end{aligned}
\end{equation}
where ${\alpha }_{j}={{\alpha }_{0}}{{\rho }_{j}}$ with ${{\alpha }_{0}}$ and ${{\rho }_{j}}\sim \mathcal{C}\mathcal{N}\left( 0,\sigma _{{\rho },j}^{2} \right)$ denoting the complex amplitude of the target and the $j$-th patch, respectively; ${\textbf{b}_{t}}\left( \theta  \right)$ and ${\textbf{b}_{r}}\left( \theta  \right)$ denote the transmit steering vector and receive steering vector of the radar at angle $\theta$, respectively, given by
\begin{equation}\nonumber
{\textbf{b}_{t}}\left( \theta  \right)=\frac{1}{\sqrt{{{M}_{T}}}}{{\left[ 1,{{e}^{-j\pi \sin \theta }},\ldots ,{{e}^{-j\pi \left( {{M}_{T}}-1 \right)\sin \theta }} \right]}^{T}},
\end{equation}
\begin{equation}\nonumber
{\textbf{b}_{r}}\left( \theta  \right)=\frac{1}{\sqrt{{{M}_{R}}}}{{\left[ 1,{{e}^{-j\pi \sin \theta }},\ldots ,{{e}^{-j\pi \left( {{M}_{R}}-1 \right)\sin \theta }} \right]}^{T}}.
\end{equation}

Assume that the clutter region contains $Q$ elements causing clutter return, in which the $q$-th element is located in the angular direction of ${{\tilde{\theta }}_{q}}$. Then, the clutter return, denoted by ${\textbf{Y}_{c}}$, can be given by
\begin{equation}
{\textbf{Y}_{c}}=\sum\limits_{q=1}^{Q}{{{{\tilde{\alpha }}}_{q}}{\textbf{b}_{r}}\big( {{{\tilde{\theta }}}_{q}} \big)\textbf{b}_{t}^{T}\big( {{{\tilde{\theta }}}_{q}} \big)\textbf{S}\textbf{J}_{{{\tilde{\tau }}}_{q}} },
\end{equation}
where ${{\tilde{\alpha }}_{q}}\sim \mathcal{C}\mathcal{N}\left( 0,\sigma _{\tilde{\alpha },q}^{2} \right)$ accounts for the complex scattering coefficient of the $q$-th element from clutter region; ${\tilde{\tau}}_q$ denotes the relative delay compared with the target echo along the LoS; and ${\textbf{J}_{k}}\in {{\mathbb{C}}^{K\times K}}$ is the shift matrix with
\begin{equation}\nonumber
	{\textbf{J}_{k}}\left( i,j \right)=\left\{ \begin{matrix}
		1,\text{ }\text{if}\text{ }i-j=k  \\
		0,\text{ }\text{if}\text{ }i-j\neq k  \\
	\end{matrix} \right..
\end{equation}

We also consider $G$-path channels between the BS and the radar receiver. Denote the {direction of departure} (DoD) and direction of arrival (DoA) of the $g$-th path by $\varphi _{cr}^{g}$ and $\phi _{cr}^{g}$, respectively. Then, the interference signal of the communication system on the radar system, denoted by ${\textbf{Y}_{cr}}$, can be modeled by
\begin{equation}
{\textbf{Y}_{cr}}=\sum\limits_{g=1}^{G}{{{\beta }_{g}}{\textbf{b}_{r}}\left( \phi _{cr}^{g} \right)\textbf{a}_{t}^{T}\left( \varphi _{cr}^{g} \right)\textbf{X}_{g}},
\end{equation}
where ${{\beta }_{g}}\sim \mathcal{C}\mathcal{N}\big( 0,\sigma _{\beta ,g}^{2} \big)$ is the gain of the $g$-th path; ${\textbf{a}_{t}}\left( \varphi  \right)$ denotes the transmit steering vector of the BS at angle $\varphi $, given by
\begin{equation}\nonumber
{\textbf{a}_{t}}\left( \varphi  \right)=\frac{1}{\sqrt{{{N}_{T}}}}{{\left[ 1,{{e}^{-j\pi \sin \varphi}},\ldots ,{{e}^{-j\pi \left( {{N}_{T}}-1 \right)\sin \varphi }} \right]}^{T}};
\end{equation}
and $\textbf{X}_{g}\in {{\mathbb{C}}^{N_T\times K}}$ denotes the communication data segment received by the radar along the $g$-th path.

Finally, the received space-time signal, denoted by $\textbf{Y}\in {{\mathbb{C}}^{{{M}_{R}}\times {K}}}$, can be modeled as
\begin{equation}
	\textbf{Y}={\textbf{Y}_{m}}+{\textbf{Y}_{c}}+{\textbf{Y}_{cr}}+{\textbf{Y}_{n}},
\end{equation}
where ${\textbf{Y}_{n}}$ denotes the additive noise, the entries of which are modeled as independent complex Gaussian random variables with zero mean and variance $\sigma _{r}^{2}$. We can recast $\textbf{Y}$ in vector form as
\begin{equation}
\begin{aligned}
\textbf{y}=\sum\limits_{j=0}^{J}{{{\alpha }_{j}}{\textbf{H}_{j}}\textbf{s}}+\sum\limits_{q=1}^{Q}{{{{\tilde{\alpha }}}_{q}}{{{\tilde{\textbf{H}}}}_{q}}\textbf{s}}+\sum\limits_{g=1}^{G}{{{\beta }_{g}}\textbf{H}_{cr}^{g}\textbf{x}_{g}}+{\textbf{y}_{n}},
\end{aligned}
\end{equation}
where ${\textbf{H}_{0}}={\textbf{I}_{K}}\otimes \left( {\textbf{b}_{r}}\left( {{\theta }_{0}} \right)\textbf{b}_{t}^{T}\left( {{\theta }_{0}} \right) \right)$; ${\textbf{H}_{j}}={\textbf{I}_{K}}\otimes \left( {\textbf{b}_{r}}\left( {{\theta }_{0}} \right)\textbf{b}_{t}^{T}\left( {{\theta }_{j}} \right)+{\textbf{b}_{r}}\left( {{\theta }_{j}} \right)\textbf{b}_{t}^{T}\left( {{\theta }_{0}} \right) \right)$; ${{\tilde{\textbf{H}}}_{q}}={\textbf{J}^{T}_{{{\tilde{\tau }}}_{q}} }\otimes \big( {\textbf{b}_{r}}\big( {{{\tilde{\theta }}}_{q}} \big)\textbf{b}_{t}^{T}\big( {{{\tilde{\theta }}}_{q}} \big) \big)$; $\textbf{H}_{cr}^{g}={\textbf{I}_{K}}\otimes \left({\textbf{b}_{r}}\left( \phi _{cr}^{g} \right)\textbf{a}_{t}^{T}\left( \varphi _{cr}^{g} \right)\right)$; $\textbf{s}=\text{vec}\left( \textbf{S} \right)$; ${\textbf{x}_{g}}={\rm vec}\left( \textbf{X}_{g} \right)$ and ${\textbf{y}_{n}}={\rm vec}\left( {\textbf{Y}_{n}} \right)$. 

\newcounter{mytempeqncnt1}
\begin{figure*}[htb]
	\normalsize
	\setcounter{mytempeqncnt1}{\value{equation}}
	\setcounter{equation}{5}
	\begin{equation}
	\text{SINR}\left( \textbf{w}, \textbf{s},\textbf{V} \right)=\frac{\mathbb{E}\left\{ {{\left| {\textbf{w}^{H}}\sum\limits_{j=0}^{J}{{{\alpha }_{j}}{\textbf{H}_{j}}\textbf{s}} \right|}^{2}} \right\}}{\mathbb{E}\left\{ {{\left| {\textbf{w}^{H}}\sum\limits_{q=1}^{Q}{{{{\tilde{\alpha }}}_{q}}{{{\tilde{\textbf{H}}}}_{q}}\textbf{s}} \right|}^{2}} \right\}+\mathbb{E}\left\{ {{\left| {\textbf{w}^{H}}\sum\limits_{g=1}^{G}{{{\beta }_{g}}\textbf{H}_{cr}^{g}{\textbf{x}_{g}}} \right|}^{2}} \right\}+\sigma _{r}^{2}{\textbf{w}^{H}}\textbf{w}} =\sum\limits_{j=0}^{J}{\sigma _{\alpha ,j}^{2}\frac{{\textbf{w}^{H}}{\textbf{H}_{j}}\textbf{s}{\textbf{s}^{H}}\textbf{H}_{j}^{H}\textbf{w}}{{\textbf{w}^{H}}\textbf{R}\left( \textbf{V}, \textbf{s} \right)\textbf{w}}}. \label{SINR}
	\end{equation}
	\setcounter{equation}{\value{mytempeqncnt1}}
	\hrulefill
\end{figure*}
\setcounter{equation}{6}

The SINR can be used to measure the detection performance of the radar system \cite{chen2022generalized,chen2023joint,chen2022full}. In order to achieve the best detection performance, we combine the multi-path returns to maximize the output SINR\footnote{The detection probability monotonically increases with respect to (w.r.t.) the output SINR under Gaussian conditions \cite{4626391}.}. After \textbf{y} being filtered by a space-time filter $\textbf{w}\in {{\mathbb{C}}^{{{M}_{R}}{K}}}$, the SINR can be constructed as (\ref{SINR}), as shown at the top of next page, where $\sigma _{\alpha ,0}^{2}={{\left| {{\alpha }_{0}} \right|}^{2}}$ and $\sigma _{\alpha ,j}^{2}={{\left| {{\alpha }_{0}} \right|}^{2}}\sigma _{\rho ,j}^{2},j=1,\ldots,J$ and
\begin{equation}
\begin{aligned}
\textbf{R}\left( \textbf{V}, \textbf{s} \right)=&\sum\limits_{g=1}^{G}{\sigma _{\beta ,g}^{2}\textbf{H}_{cr}^{g}\left( {\textbf{I}_{{K}}}\otimes\textbf{V}{\textbf{V}^{H}} \right)}{{\left( \textbf{H}_{cr}^{g} \right)}^{H}}+\\
&\sum\limits_{q=1}^{Q}{\sigma _{\tilde{\alpha },q}^{2}{{{\tilde{\textbf{H}}}}_{q}}\textbf{s}{\textbf{s}^{H}}\tilde{\textbf{H}}_{q}^{H}}+\sigma _{r}^{2}{\textbf{I}_{{{M}_{R}}{K}}}.
\end{aligned}
\end{equation}
It can be observed from (\ref{SINR}) that the optimal $\textbf{w}$ for the space-time filter can be obtained by maximizing the generalized Rayleigh quotient of $\boldsymbol{\Psi} \left( \textbf{s} \right)=\sum\nolimits_{j=0}^{J}{\sigma _{\alpha ,j}^{2}{\textbf{H}_{j}}\textbf{s}{\textbf{s}^{H}}\textbf{H}_{j}^{H}}$ and ${\textbf{R}}\left( \textbf{V},\textbf{s} \right)$, i.e.,
\begin{equation}
\begin{aligned}
{\textbf{w}}^{\star}=\text{arg }\underset{\textbf{w}}{\mathop{\text{max}}}\frac{{\textbf{w}^{H}}\boldsymbol{\Psi} \left( \textbf{s} \right)\textbf{w}}{{\textbf{w}^{H}}\textbf{R}\left( \textbf{V},\textbf{s} \right)\textbf{w}}=\mathcal{M}\left( {\textbf{R}^{-1}}\left( \textbf{V},\textbf{s} \right)\boldsymbol{\Psi} \left( \textbf{s} \right) \right).\label{w_update}
\end{aligned}
\end{equation}
Then, using (\ref{w_update}), $\text{SINR}\left( \textbf{w}, \textbf{s},\textbf{V} \right)$ can be reformulated as
\begin{equation}
\text{SINR}\left(\textbf{s},\textbf{V} \right)=\sum\limits_{j=0}^{J}{\sigma _{\alpha ,j}^{2}}\frac{{\textbf{s}^{H}}\textbf{H}_{j}^{H}\textbf{w}{\textbf{w}^{H}}{\textbf{H}_{j}}\textbf{s}}{{\textbf{s}^{H}}\tilde{\textbf{R}}\textbf{s}+{r}\left( \textbf{V} \right)},\label{radar_obj}
\end{equation}
where ${r}\left( \textbf{V} \right)={\textbf{w}^{H}}\big( \sum\nolimits_{g=1}^{G}{\sigma _{\beta ,g}^{2}\textbf{H}_{cr}^{g}}\left( {\textbf{I}_{{K}}}\otimes \textbf{V}{\textbf{V}^{H}} \right){{\left( \textbf{H}_{cr}^{g} \right)}^{H}}+\sigma _{r}^{2}{\textbf{I}_{{{M}_{R}}{{K}}}} \big)\textbf{w}$ and $\tilde{\textbf{R}}=\sum\nolimits_{q=1}^{Q}{{\sigma _{\tilde{\alpha },q}^{2}}\tilde{\textbf{H}}_{q}^{H}\textbf{w}{\textbf{w}^{H}}{{{\tilde{\textbf{H}}}}_{q}}}$. In the special case of $J=0$, i.e., the radar LoS detection, $\text{SINR}\left( \textbf{w}, \textbf{s},\textbf{V} \right)$ can be rewritten as
\begin{equation}\nonumber
\text{SINR}\left(\textbf{s},\textbf{V} \right)={\sigma _{\alpha ,0}^{2}}{\textbf{s}^{H}}\textbf{H}_{0}^{H}{\textbf{R}}^{-1}\left( \textbf{V}, \textbf{s} \right){\textbf{H}_{0}}\textbf{s},
\end{equation}
which has the same form as that in \cite{tang2016joint,wu2017transmit}.

We assume that ${\sigma _{\alpha ,0}^{2}}$ is known, $\left\{ {{\theta }_{j}},{\sigma _{\rho ,j}^{2}} \right\}_{j=1}^{J}$, $\big\{ {{{\tilde{\theta }}}_{q}},{\sigma _{\tilde{\alpha },q}^{2}} \big\}_{q=1}^{Q}$ and $\big\{ \varphi _{cr}^{g},\phi _{cr}^{g},{\sigma _{\beta ,g}^{2}} \big\}_{g=1}^{G}$ can be acquired by cognitive paradigm \cite{7470514,8477186,8352726,9582836,6404093}.

\subsection{Communication Performance}
The signal received by the CU in the $p$-th PRI is subject to intermittent radar interference and noise. Then, the received signal at time instant $k$ can be modeled by
\begin{equation}
\textbf{r}\left( k \right)={\textbf{r}_{m}}\left( k \right)+{\textbf{r}_{rc}}\left( k \right)+{\textbf{r}_{n}}\left( k \right),\label{CU_receive}
\end{equation}
where ${\textbf{r}_{m}}\left( k \right)$ denotes the desired communication multi-path signal, ${\textbf{r}_{rc}}\left( k \right)$ denotes the interference of the radar system on the communication system, and ${\textbf{r}_{n}}\left( k \right)$ denotes the additive noise. The entries of ${\textbf{r}_{n}}\left( k \right)$ are independent complex Gaussian random variables with zero mean and variance $\sigma _{c}^{2}$.

Denote $L$ as the number of paths between the BS and the CU. The DoD and DoA of the $l$-th  path are represented by $\vartheta _{t}^{l}$ and $\vartheta _{r}^{l}$, respectively. Since the delay of each path will distort the received communication symbols at the CU, we assume that the effect can be perfectly pre-compensated at the BS with given estimated delay parameters by using existing compensation methods \cite{7343329,7504275}. Then, ${\textbf{r}_{m}}\left( k \right)$ can be formulated as
\begin{equation}
{\textbf{r}_{m}}\left( k \right)=\sum\limits_{l=1}^{L}{{{\upsilon }_{l}}{\textbf{G}_{l}}\textbf{x}\left( k \right)},\label{CU_receive_com}
\end{equation}
where ${{\upsilon}_{l}}\sim \mathcal{C}\mathcal{N}\big( 0,\sigma _{\upsilon ,l}^{2} \big)$ and ${\textbf{G}_{l}}={\textbf{a}_{r}}\left( \vartheta _{r}^{l} \right)\textbf{a}_{t}^{T}\left( \vartheta _{t}^{l} \right)$ denote the gain and the transmit-receive steering matrix of the $l$-th path, respectively, while ${\textbf{a}_{r}}\left( \vartheta  \right)$ is the receive steering vector of the CU at angle $\vartheta $, given by
\begin{equation}\nonumber
{\textbf{a}_{r}}\left( \vartheta \right)=\frac{1}{\sqrt{{{N}_{R}}}}{{\left[ 1,{{e}^{-j\pi \sin \vartheta}},\ldots ,{{e}^{-j\pi \left( {{N}_{R}}-1 \right)\sin \vartheta }} \right]}^{T}}.
\end{equation}

Assuming $I$ scatterers reflecting the radar pulses towards the CU, ${\textbf{r}_{rc}}\left( k \right)$ can be expressed as
\begin{equation}
{\textbf{r}_{rc}}\left( k \right)=\sum\limits_{i=1}^{I}{{{\gamma }_{i}}{\textbf{a}_{r}}\left( \varphi _{rc}^{i} \right)\textbf{b}_{t}^{T}\left( \phi _{rc}^{i} \right)\textbf{s}\left( k-{\tau}_{rc}^{i} \right)},
\end{equation}
where ${{\gamma}_{i}}\sim \mathcal{C}\mathcal{N}\big( 0,\sigma _{\gamma ,i}^{2} \big)$ accounts for the gain of the $i$-th scattering path; $\varphi _{rc}^{i}$, $\phi _{rc}^{i}$ and ${\tau}_{rc}^{i}\in \big\{0,\ldots ,\tilde{K}-1 \big\}$ are the DoA, DoD and the delay of the $i$-th path, respectively. The intermittent nature of the radar interference can be characterized by the following matrix ${\textbf{L}_{i}}$ \cite{9582836}
\begin{equation}\nonumber
{\textbf{L}_{i}}=\bigg\{ 
\begin{matrix}
\begin{aligned}
&\big[\begin{matrix}
{\textbf{0}_{K,{\tau}_{rc}^{i}}} & {\textbf{I}_{K}} & {\textbf{0}_{K,\tilde{K}-K-{\tau}_{rc}^{i}}}  
\end{matrix}\big],\qquad\quad\text{ if }{\tau}_{rc}^{i}\in \mathcal{K}1  \\
&\big[\begin{matrix}
{\textbf{J}_{\tilde{K}-{\tau}_{rc}^{i}}} & {\textbf{0}_{K,\tilde{K}-2K}} & {\textbf{J}_{ -K+\tilde{K}-{\tau}_{rc}^{i} }}  
\end{matrix} \big],\text{ if }{\tau}_{rc}^{i}\in \mathcal{K}2  \\
\end{aligned}
\end{matrix},
\end{equation}
where $\mathcal{K}1=\big\{ 0,\ldots ,\tilde{K}-K \big\} $, $\mathcal{K}2=\big\{ \tilde{K}-K+1,\ldots ,\tilde{K}-1 \big\}$. Then, we have $
\textbf{s}\left( k-{\tau} _{rc}^{i} \right)=\textbf{S}{\textbf{L}_{i}}\textbf{e}\big( k-\big( p-1 \big)\tilde{K} \big)
=\left( {\textbf{e}^{T}}\big( k-\left( p-1 \right)\tilde{K} \big){\textbf{L}_{i}^{T}}\otimes {\textbf{I}_{{{M}_{T}}}} \right)\textbf{s},$
where $\textbf{e}\left( n \right)\in {{\mathbb{C}}^{{\tilde{K}}}}$ is a direction vector whose $n$-th entry is one, while all the other entries equal to zero. Thus, the radar interference ${\textbf{r}_{rc}}\left( k \right)$ can be rewritten as
\begin{equation}
{\textbf{r}_{rc}}\left( k \right)=\sum\limits_{i=1}^{I}{{{\gamma }_{i}}\textbf{G}_{rc}^{i}\left( k \right)\textbf{s}},\label{CU_receive_radar}
\end{equation}
where $\textbf{G}_{rc}^{i}\left( k \right)={\textbf{a}_{r}}\left( \varphi _{rc}^{i} \right)\textbf{b}_{t}^{T}\left( \phi _{rc}^{i} \right)\big( {\textbf{e}^{T}}\big( k-\left( p-1 \right)\tilde{K} \big){{ {\textbf{L}_{i}^{T}} }}\otimes {\textbf{I}_{{{M}_{T}}}} \big)$.

Using (\ref{CU_receive}), (\ref{CU_receive_com}) and (\ref{CU_receive_radar}), the achievable transmission rate at time instant $k$ can be calculated as
\begin{equation}
{{\rm MI}_{k}}\left( \textbf{s},\textbf{V} \right)=\log \left| \left( \sum\limits_{l=1}^{L}{\sigma _{\upsilon ,l}^{2}\textbf{G}_{l}\textbf{V}{\textbf{V}^{H}}{\textbf{G}_{l}^{H}}} \right){{\left( \textbf{R}_{c}^{k}\left( \textbf{s} \right) \right)}^{-1}}+{\textbf{I}_{{{N}_{R}}}} \right|,\label{MI_k}
\end{equation}
where $\textbf{R}_{c}^{k}\left( \textbf{s} \right)=
\sigma _{c}^{2}{\textbf{I}_{{{N}_{R}}}}+\sum\nolimits_{i=1}^{I}{\sigma _{\gamma ,i}^{2}\textbf{G}_{rc}^{i}\left( k \right)\textbf{s}{\textbf{s}^{H}}{{\left( \textbf{G}_{rc}^{i}\left( k \right) \right)}^{H}}}$. Then, the average communication rate during the $p$-th PRI is given by
\begin{equation}
{\rm MI}\left( \textbf{s},\textbf{V} \right)=\frac{1}{{\tilde{K}}}\sum\limits_{k=1}^{{\tilde{K}}}{{{\rm MI}_{k}}\left( \textbf{s},\textbf{V} \right)}.\label{MI_sV}
\end{equation}
In the special case of $L=1$, ${\rm MI}\left( \textbf{s},\textbf{V} \right)$ can be rewritten as
\begin{equation}\nonumber
	{\rm MI}\left( \textbf{s},\textbf{V} \right)=\frac{1}{{\tilde{K}}}\sum\limits_{k=1}^{{\tilde{K}}}{\log \left|  {\sigma _{\upsilon ,l}^{2}\textbf{G}_{l}\textbf{V}{\textbf{V}^{H}}{\textbf{G}_{l}^{H}}} {{\left( \textbf{R}_{c}^{k}\left( \textbf{s} \right) \right)}^{-1}}+{\textbf{I}_{{{N}_{R}}}} \right|},
\end{equation}
which has a similar form as that in \cite{7470514,9582836}.

Similar to the radar system, we also assume that $\big\{{\textbf{G}_{l}},\sigma_{\upsilon ,l}^{2} \big\}_{l=1}^{L}$ and $\left\{\varphi _{rc}^{i},\phi _{rc}^{i},\sigma_{\gamma ,i}^{2} \right\}_{i=1}^{I}$ can be estimated. {We assume that MIMO radar and MIMO communication systems are coordinated through a fusion center \cite{8352726,8477186}. The fusion center can collect channel state information for each system, design appropriate transmit waveforms, and assign them to the corresponding systems.}

\subsection{Problem Formulation}
For the radar system, the relevant figure of merit is the output SINR defined in (\ref{radar_obj}). {We can see that the communication interference brings challenges to the radar system to combine the target return along the LoS path and $J$ NLoS paths.} For the communication system, the relevant merit is the transmission rate given by (\ref{MI_sV}). The radar system presents different interference covariance matrix $\textbf{R}_{c}^{k}\left( \textbf{s} \right)$ at each time instant of communication data transmission. This non-homogeneous interference increases the difficulty of $L$ communication paths combining. To effectively suppress the mutual interference, we jointly design the radar waveform $\textbf{s}$ and the communication precoder $\textbf{V}$ to maximize the radar SINR while guaranteeing a minimum required communication transmission rate, denoted by ${{\rm MI}_{0}}$. The corresponding optimization problem is formulated as follows.
\begin{subequations}
\begin{align}
\mathcal{P}\text{0: }&\underset{\textbf{s},\textbf{V}}{\mathop{\max}}\text{ }
\sum\limits_{j=0}^{J}{\sigma _{\alpha ,j}^{2}}\frac{{\textbf{s}^{H}}\textbf{H}_{j}^{H}\textbf{w}{\textbf{w}^{H}}{\textbf{H}_{j}}\textbf{s}}{{\textbf{s}^{H}}\tilde{\textbf{R}}\textbf{s}+{r}\left( \textbf{V} \right)}\notag \\ 
&\text{s.t.}\quad{\rm MI}\left( \textbf{s},\textbf{V} \right)\ge {{\rm MI}_{0}}\label{com_MI} \\ 
&\qquad\text{ }{\rm tr}\left( \textbf{V}{\textbf{V}^{H}} \right)\le {{P}_{B}}\label{com_power}\\
&\qquad\text{ }{{\left\| \textbf{s} \right\|}^{2}}\le {{P}_{R}},\label{radar_power}
\end{align}
\end{subequations}
where ${{P}_{B}}$ and ${{P}_{R}}$ denote the maximum transmit power of the BS and the radar, respectively.

Before proceeding to solve $\mathcal{P}\text{0}$, we compare our formulated problem with those of the existing works related to radar-centric CRC. In \cite{7470514}, the authors study the co-design of the MIMO-MC radar sampling scheme and the communication covariance matrix to reduce mutual interference. However, the proposed communication design cannot generalize to the multi-path scenarios we consider. In \cite{8352726,9582836}, the radar waveform, receive filter and communication codebook are jointly designed to achieve the tradeoff between radar SINR and communication rate. However, the multi-path combining of useful signals for each system is not considered and the proposed communication codebook and radar waveform design algorithms don't work for $\mathcal{P}\text{0}$. In our constructed problem, the radar SINR and the communication rate have more complex forms than those in previous works, which makes it more difficult to deal with the non-concave objective function and non-convex constraint (\ref{com_MI}). We can adopt the alternating optimization method to decouple $\textbf{V}$ and $\textbf{s}$. Specifically, we will decompose $\mathcal{P}\text{0}$ into two sub-problems for $\textbf{V}$ and $\textbf{s}$, respectively. Multi-path propagation becomes a challenge to solving each non-convex sub-problem. We will then develop fast algorithms with a polynomial computational complexity to get high-quality solutions.

\section{Communication Precoder Design}
In this section, we will consider the optimization of the communication precoder $\textbf{V}$ with fixed radar waveform $\textbf{s}$, that is,
\begin{equation}
\begin{aligned}
\mathcal{P}\text{1: }&\underset{\textbf{V}}{\mathop{\min}}\text{ }r\left( \textbf{V} \right)\text{ } \\ 
&\text{s.t.}\quad(\rm \ref{com_MI}), (\rm \ref{com_power}).
\end{aligned}
\end{equation}
$\mathcal{P}\text{1}$ is challenging due to the non-convex constraint (\ref{com_MI}). When $L=1$, $\mathcal{P}\text{1}$ has a similar form to the communication space-time covariance matrix design in \cite{7470514,8352726}. Using the Lagrange dual-decomposition method and ellipsoid method can effectively design the communication precoder in this special scenario. When $L>1$, multi-path propagation and the dynamic interference make it difficult to deal with the non-convexity. In the following, we will obtain a general precoder by solving $\mathcal{P}\text{1}$ with the SCA and ADMM.

\subsection{Sub-Optimal Solution Based on SCA}
This subsection focuses on approximating $\mathcal{P}\text{1}$ by a series of convex QCQP problems. We first need to construct a surrogate function to approximate ${\rm MI}\left( \textbf{s},\textbf{V} \right)$ defined in (\ref{MI_sV}), so that the SCA method can be used to eliminate the non-convexity of (\ref{com_MI}). Since ${\rm MI}\left( \textbf{s},\textbf{V} \right)$ is non-convex w.r.t. $\textbf{V}$, it is challenging to derive its minorizer using the first-order Taylor expansion. To proceed, we resort to the following lemma to convert ${\rm MI}\left( \textbf{s},\textbf{V} \right)$ to a tractable form.

\emph{\textbf{Lemma 1} (Transformation of} ${\rm MI}\left( \textbf{s},\textbf{V} \right)$\emph{): } ${\rm MI}\left( \textbf{s},\textbf{V} \right)$ can be equivalently rewritten as
\begin{equation}
{\rm MI}\left( \textbf{v} \right)=\frac{1}{{\tilde{K}}}\sum\limits_{k=1}^{\tilde{K}}{\log \left| \textbf{C}{{\left( {\textbf{E}_{k}}\left( \textbf{v} \right) \right)}^{-1}}{\textbf{C}^{H}} \right|}\label{new_MI},
\end{equation}
where $\textbf{v}={\rm vec}\left( \textbf{V} \right)$, $\textbf{C}=\left[ {\textbf{I}_{{{N}_{R}}{{N}_{T}}D}},{\textbf{0}_{{{N}_{R}}{{N}_{T}}D\times {{N}_{R}}}} \right]$ and
\begin{equation}
\begin{aligned}
&{\textbf{E}_{k}}\left( \textbf{v} \right)\\
&=\left[ \begin{matrix}
{\textbf{I}_{{{N}_{R}}{{N}_{T}}D}} & {{\boldsymbol{\Delta} }^{\frac{1}{2}}}\left( {\textbf{I}_{{{N}_{R}}}}\otimes {\textbf{v}^{*}} \right)  \\
\left( {\textbf{I}_{{{N}_{R}}}}\otimes {\textbf{v}^{T}} \right){{\boldsymbol{\Delta} }^{\frac{1}{2}}} & \textbf{R}_{c}^{k}\left( \textbf{s} \right)+\left( {\textbf{I}_{{{N}_{R}}}}\otimes {\textbf{v}^{T}} \right)\boldsymbol{\Delta} \left( {\textbf{I}_{{{N}_{R}}}}\otimes {\textbf{v}^{*}} \right)  \\
\end{matrix} \right].\label{E_k}
\end{aligned}
\end{equation}
with ${\boldsymbol{\Delta}}$ defined in (\ref{delta}).
\begin{IEEEproof}
See Appendix A.	
\end{IEEEproof}

Note that ${\rm MI}\left( \textbf{v} \right)$ is convex w.r.t. ${\textbf{E}_{k}}\left( \textbf{v} \right)$ \cite{7895135}. Then, we can use its first-order condition to approximate (\rm \ref{com_MI}) by a series of convex constraints given as follows.

\emph{\textbf{Lemma 2} (SCA-Based Transformation of Constraint} (\rm \ref{com_MI})\emph{): } (\rm \ref{com_MI}) can be successively approximated by the following convex quadratic constraint
\begin{equation}
{\textbf{v}^{H}}{{\bar{\boldsymbol{\Gamma}}}_{22}}\left( {\bar{\textbf{v}}} \right)\textbf{v}-2\mathcal{R}\left( {{{\bar{\boldsymbol{\Gamma}}}}_{12}}\left( {\bar{\textbf{v}}} \right)\textbf{v} \right)\le -\overline{\rm MI}\left( {\bar{\textbf{v}}} \right)\label{com_sca},
\end{equation}
where ${\bar{\boldsymbol{\Gamma}}_{22}\left( {\bar{\textbf{v}}} \right)}$, ${\bar{\boldsymbol{\Gamma}}}_{12}\left( {\bar{\textbf{v}}} \right)$ and $\overline{\rm MI}\left( {\bar{\textbf{v}}} \right)$ are defined in (\ref{tao_22}), (\ref{tao_12}) and (\ref{MI_bar}), respectively, with $\bar{\textbf{v}}$ being the solution in the previous SCA iteration.
\begin{IEEEproof}
See Appendix B.	
\end{IEEEproof}

Similar to constraint (\ref{com_sca}), we also need to convert the objective function $r\left(\textbf{V}\right)$ into a convex one w.r.t. \textbf{v}. Before doing so, we first define $\textbf{W}=\textbf{w}{\textbf{w}^{H}}$ and partition it into a block matrix, i.e., $\textbf{W}={{\left( {\textbf{W}_{ij}} \right)}_{{{K}}\times {{K}}}}$, where ${\textbf{W}_{ij}}\in {{\mathbb{C}}^{{{M}_{R}}\times {{M}_{R}}}}$ can be calculated by
${\textbf{W}_{ij}}=\textbf{w}\left( \left( i-1 \right){{M}_{R}}+1:i{{M}_{R}} \right){\textbf{w}^{H}}\left( \left( j-1 \right){{M}_{R}}+1:j{{M}_{R}} \right)$ for $\big( i,j \big)\in {{\big\{ 1,\ldots ,{K} \big\}}^{2}}$. Then, $r\left(\textbf{V} \right)$ can be transformed to a convex quadratic function in the following lemma.}

\emph{\textbf{Lemma 3} (Transformation of} $r\left(\textbf{V} \right)$\emph{):} $r\left(\textbf{V} \right)$ can be equivalently rewritten as
\begin{equation}
\tilde{r}\left( \textbf{v} \right)={\textbf{v}^{H}}{\boldsymbol{\Pi}}\textbf{v},\label{objective}
\end{equation}
where $
{\boldsymbol{\Pi} }={\textbf{I}_{D}}\otimes \left( \sum\nolimits_{g=1}^{G}{\sum\nolimits_{i=1}^{{K}}{\sigma _{\beta ,g}^{2}{\left(\textbf{T}_{cr}^{g}\right)}^{H}{\textbf{W}_{ii}}\textbf{T}_{cr}^{g}}} \right)
$
with
${\textbf{T}}_{cr}^{g}={\textbf{b}_{r}}\left( \phi _{cr}^{g} \right)\textbf{a}_{t}^{T}\left( \varphi _{cr}^{g} \right)$.
\begin{IEEEproof}
After removing the term unrelated to $\textbf{V}$, ${r}\left(\textbf{V} \right)$ can be rewritten as	
	\begin{equation}
	\begin{aligned}
	\tilde{r}\left( \textbf{V} \right)&={\textbf{w}^{H}}\left( \sum\limits_{g=1}^{G}{\sigma_{\beta,g}^{2}\textbf{H}_{cr}^{g}\left( {\textbf{I}_{{{K}}}}\otimes \textbf{V}{\textbf{V}^{H}} \right){{\big( \textbf{H}_{cr}^{g} \big)}^{H}}} \right)\textbf{w} \\ 
	& \overset{(\rm a)}{=}{\textbf{w}^{H}}\left( \sum\limits_{g=1}^{G}{\sigma _{\beta ,g}^{2}{\textbf{I}_{{{K}}}}\otimes \left( \textbf{T}_{cr}^{g}\textbf{V}{\textbf{V}^{H}}{\big(\textbf{T}_{cr}^{g}\big)}^{H}\right)} \right)\textbf{w} \\ 
	&=\sum\limits_{g=1}^{G}{\sigma _{\beta ,g}^{2}{\rm tr}\left( \left( {\textbf{I}_{{{K}}}}\otimes \left(\textbf{T}_{cr}^{g}\textbf{V}{\textbf{V}^{H}}{\big(\textbf{T}_{cr}^{g}\big)}^{H}\right) \right)\textbf{W} \right)} \\ 
	&=\sum\limits_{g=1}^{G}{\sigma _{\beta ,g}^{2}\sum\limits_{i=1}^{{{K}}}{{\rm tr}\left(\textbf{T}_{cr}^{g}\textbf{V}{\textbf{V}^{H}}
			{\big(\textbf{T}_{cr}^{g}\big)}^{H}{\textbf{W}_{ii}} \right)}} \\ 
	&={\rm tr}\left( \textbf{V}{\textbf{V}^{H}}\left( \sum\limits_{g=1}^{G}{\sum\limits_{i=1}^{{{K}}}{\sigma _{\beta ,g}^{2}{\big(\textbf{T}_{cr}^{g}\big)}^{H}{\textbf{W}_{ii}}\textbf{T}_{cr}^{g}}} \right) \right) \\ 
	&\overset{(\rm b)}{=}{\textbf{v}^{H}}{\boldsymbol{\Pi} }\textbf{v},  
	\end{aligned}
	\end{equation}
	where the procedure (a) comes from $\textbf{H}_{cr}^{g}={\textbf{I}_{{{K}}}}\otimes {\textbf{b}_{r}}\left( \phi _{cr}^{g} \right)\textbf{a}_{t}^{T}\left( \varphi _{cr}^{g} \right)$ and the procedure (b) uses the identity that ${\rm tr}\left( \textbf{A}_1\textbf{A}_2\textbf{A}_3\textbf{A}_4 \right)={{\rm vec}^{T}}\left( \textbf{A}_4 \right)\left( \textbf{A}_1\otimes {\textbf{A}_3^{T}} \right){\rm vec}\left( {\textbf{A}_2^{T}} \right)$. It is easy to verify that ${\boldsymbol{\Pi}}$ is Hermitian positive semidefinite.
\end{IEEEproof}

Combining the above lemmas, we have the following proposition.

\emph{\textbf{Proposition 1} (SCA-Based Transformation of $\mathcal{P}{\rm 1}$): }$\mathcal{P}\text{1}$ can be successively approximated by the following QCQP problem
\begin{equation}
\begin{aligned}
\mathcal{P}\text{1.1}\big( \bar{\textbf{v}}\big)\text{: }&\underset{\textbf{v}}{\mathop{\min }}\text{ }{\textbf{v}^{H}}{\boldsymbol{\Pi}}\textbf{v} \\ 
&\text{s.t.}\quad(\ref{com_sca})\\
&\qquad\text{ }{\textbf{v}^{H}}\textbf{v}\le {{P}_{B}}.
\end{aligned}
\end{equation}
Denoting the $s$-th solution of the SCA procedure as $\textbf{v}^{(s)}$, we can obtain $\textbf{v}^{(s+1)}$ by solving $\mathcal{P}\text{1.1}\big( {\textbf{v}}^{(s)}\big)$. 
\begin{IEEEproof}
With Lemma 3, we give a convex quadratic representation of original objective function $r\left( \textbf{V} \right)$. By applying Lemma 1 and Lemma 2, we convert
the non-convex constraint (${\rm \ref{com_MI}}$) into a convex one. By using the identity that ${\rm tr}\big( {\textbf{A}_1^{H}}\textbf{A}_2 \big)={{\rm vec}^{H}}\left( \textbf{A}_1 \right){\rm vec}\left( \textbf{A}_2 \right)$, we convert the constraint (\rm \ref{com_power}) to ${\textbf{v}^{H}}\textbf{v}\le {{P}_{B}}$. Thus, $\mathcal{P}\text{1.1}\big( \bar{\textbf{v}}\big)$ is a convex QCQP problem.
\end{IEEEproof}

$\mathcal{P}\text{1.1}\big( \bar{\textbf{v}}\big)$ can be cast as a second-order cone programming (SOCP) problem and solved by the {interior-point methods} (IPM) \cite{nesterov1994interior}. Solving the SOCP problem requires $\mathcal{O}\left( 2\log \left( {1}/{\varepsilon } \right) \right)$ iterations to converge, where ${\varepsilon}$ denotes the relative accuracy and each iteration has a computational complexity of $\mathcal{O}\big( {{\left( D{{N}_{T}} \right)}^{3}}+D{{N}_{T}}\big( {{\left( D{{N}_{T}} \right)}^{2}}+{{\left( D{{N}_{T}}+1 \right)}^{2}} \big) \big)$ \cite{6891348}. 
Denote $N_{s1}$ as the number of iterations for the SCA method. Then, the total complexity of solving $\mathcal{P}\text{1}$ is $\mathcal{O}\big( 2\log \left( {1}/{\varepsilon } \right){{N}_{s1}}\big( {{\left( D{{N}_{T}} \right)}^{3}}+D{{N}_{T}}\big( {{\left( D{{N}_{T}} \right)}^{2}}+{{\left( D{{N}_{T}}+1 \right)}^{2}} \big) \big) \big)$. Since the IPM will increase the computation burden in large-scale MIMO systems, we will leverage the ADMM in \cite{7517329} to more efficiently solve $\mathcal{P}\text{1.1}\big( \bar{\textbf{v}}\big)$ next.

\subsection{Low-Complexity Design through ADMM}
In this subsection, we find the optimal solution to $\mathcal{P}\text{1.1}\big( \bar{\textbf{v}}\big)$ by the ADMM. More specifically, we first transform $\mathcal{P}\text{1.1}\big( \bar{\textbf{v}}\big)$ to the following equivalent problem by introducing auxiliary variables ${\textbf{v}_{i}}\in {{\mathbb{C}}^{D{{N}_{T}}}},i=1,2$.
\begin{subequations}
\begin{align}
\mathcal{P}\text{1.2}\big( \bar{\textbf{v}}\big)\text{: }&\underset{\textbf{v},\left\{ {\textbf{v}_{i}} \right\}}{\mathop{\min}}\text{ }{\textbf{v}^{H}}\boldsymbol{\Pi} \textbf{v} \label{ADMM_objective} \\ 
&\text{s.t.}\quad\textbf{v}={\textbf{v}_{i}},i=1,2 \label{equality}\\
&\qquad\text{ }\textbf{v}_{1}^{H}{\textbf{v}_{1}}\le {{P}_{B}}\label{ADMM1} \\ 
&\qquad\text{ }\textbf{v}_{2}^{H}{{\bar{\boldsymbol{\Gamma}}}_{22}}\left( \bar{\textbf{v}} \right){\textbf{v}_{2}}-2\mathcal{R}\left( {{{\bar{\boldsymbol{\Gamma}}}}_{12}}\left( {\bar{\textbf{v}}} \right){\textbf{v}_{2}} \right)\le -\overline{\rm MI}\left( {\bar{\textbf{v}}} \right)\label{ADMM2}.
\end{align}
\end{subequations}

Let ${{\mathcal{C}}_{1}}$ and ${{\mathcal{C}}_{2}}$ denote the feasible regions of (\ref{ADMM1}) and (\ref{ADMM2}), respectively. Define the indicator functions as
\begin{equation} 
{{\mathbb{I}}_{{{\mathcal{C}}_{i}}}}\left( {\textbf{v}_{i}} \right)=\left\{ \begin{matrix}
\begin{aligned}
&0,\quad\enspace\text{ }\text{if}\text{ } {\textbf{v}_{i}}\in {{\mathcal{C}}_{i}}  \\
&+\infty,\text{otherwise}  \\
\end{aligned}
\end{matrix} \right.,i=1,2.\label{indicator}
\end{equation}
Then, using (\ref{indicator}), we can incorporate the constraints (\ref{ADMM1}) and (\ref{ADMM2}) into (\ref{ADMM_objective}) and construct the following ADMM reformulation of $\mathcal{P}\text{1.1}\big( \bar{\textbf{v}}\big)$
\begin{equation}
\begin{aligned}
\mathcal{P}\text{1.3}\big( \bar{\textbf{v}}\big)\text{: }&\underset{\textbf{v},\left\{ {\textbf{v}_{i}} \right\}}{\mathop{\min}}\text{ }{\textbf{v}^{H}}\boldsymbol{\Pi} \textbf{v}+\sum\limits_{i=1}^{2}{{{\mathbb{I}}_{{{\mathcal{C}}_{i}}}}\left( {\textbf{v}_{i}} \right)}\\ 
&\text{s.t.}\quad(\rm\ref{equality}).
\end{aligned}
\end{equation}

The augmented Lagrangian of $\mathcal{P}\text{1.3}\big( \bar{\textbf{v}}\big)$ is given by
\begin{equation}
\begin{aligned}
&\mathcal{L}\left( \textbf{v},\left\{ {\textbf{v}_{i}} \right\},\left\{ {\textbf{c}_{i}} \right\} \right)\\
&={\textbf{v}^{H}}\boldsymbol{\Pi} \textbf{v}+\sum\limits_{i=1}^{2}{{{\mathbb{I}}_{{{\mathcal{C}}_{i}}}}\left( {\textbf{v}_{i}} \right)}+\frac{{\bar{\rho}}}{2}\sum\limits_{i=1}^{2}{\left\| {\textbf{v}_{i}}-\textbf{v}+{\textbf{c}_{i}} \right\|^{2}},
\end{aligned}
\end{equation}
where ${\textbf{c}_{i}}$ is the scaled dual variable associated with the constraint $\textbf{v}={\textbf{v}_{i}}$, and $\bar{\rho }\ge 0$ denotes the penalty parameter. We can observe that $\mathcal{L}\left( \textbf{v},\left\{ {\textbf{v}_{i}} \right\},\left\{ {\textbf{c}_{i}} \right\} \right)$ can be minimized by updating $\left\{ {\textbf{v}_{i}} \right\}$ and $\textbf{v}$, alternatively. The detailed steps are listed as follows.

\emph{1)} ${\rm{\textbf{v}}_{1}}$ \emph{update:} The optimization problem w.r.t. ${\rm{\textbf{v}}_{1}}$ can be expressed as
\begin{equation}
\begin{aligned}
&\underset{{\textbf{v}_{1}}}{\mathop{\min}}\text{ }{{\left\| {\textbf{v}_{1}}-\left( \textbf{v}-{\textbf{c}_{1}} \right) \right\|}_{2}}\\ 
&\text{s.t.}\quad(\rm \ref{ADMM1}).
\label{v1_problem}
\end{aligned}
\end{equation} 
Its closed-form solution is given by
\begin{equation}\label{v1}
{\textbf{v}_{1}}=\min \left\{ \frac{\sqrt{{{P}_{B}}}}{{{\left\| \textbf{v}-{\textbf{c}_{1}} \right\|}_{2}}},1 \right\}\left( \textbf{v}-{\textbf{c}_{1}} \right).
\end{equation}

\emph{2)} ${\rm{\textbf{v}}_{2}}$ \emph{update:} The optimization problem w.r.t. ${\rm{\textbf{v}}_{2}}$ can be expressed as
\begin{equation}
\begin{aligned}
&\underset{{\textbf{v}_{2}}}{\mathop{\min}}\text{ }{{\left\| {\textbf{v}_{2}}-\left( \textbf{v}-{\textbf{c}_{2}} \right) \right\|}^{2}}\\
&\text{s.t.}\quad(\rm \ref{ADMM2}).\label{v2_problem}
\end{aligned}
\end{equation}
Using eigen-decomposition, ${{\bar{\boldsymbol{\Gamma}}}_{22}}\left( {\bar{\textbf{v}}}\right)$ becomes ${{\bar{\boldsymbol{\Gamma}}}_{22}}\left( {\bar{\textbf{v}}} \right)=\textbf{Q}\tilde{\boldsymbol{\Lambda} }{\textbf{Q}^{H}}$. Define ${{\tilde{\textbf{v}}}_{2}}={\textbf{Q}^{H}}{\textbf{v}_{2}}$, $\tilde{\textbf{t}}={\textbf{Q}^{H}}\left( \textbf{v}-{\textbf{c}_{2}} \right)$ and ${{\tilde{\boldsymbol{\Gamma} }}_{12}}\left( {\bar{\textbf{v}}} \right)={{\bar{\boldsymbol{\Gamma} }}_{12}}\left( {\bar{\textbf{v}}} \right)\textbf{Q}$. Then, we can rewrite problem $(\ref{v2_problem})$ as 
\begin{subequations}
\begin{align}
&\underset{{{{\tilde{\textbf{v}}}}_{2}}}{\mathop{\min }}\text{ }{{\left\| {{{\tilde{\textbf{v}}}}_{2}}-\tilde{\textbf{t}} \right\|}^{2}}\\
&\text{s.t.}\quad\tilde{\textbf{v}}_{2}^{H}\tilde{\boldsymbol{\Lambda} }{{\tilde{\textbf{v}}}_{2}}-2\mathcal{R}\left( {{{\tilde{\boldsymbol{\Gamma} }}}_{12}}\left( {\bar{\textbf{v}}} \right){{{\tilde{\textbf{v}}}}_{2}} \right)\le -\overline{\rm MI}\left( {\bar{\textbf{v}}} \right).\label{new_v2_constraint}
\end{align}\label{new_v2_problem}\vspace{-12pt}
\end{subequations}

Setting the gradient of the Lagrangian of problem (\ref{new_v2_problem}) to zero, we derive the optimal solution as
\begin{equation}
{{\tilde{\textbf{v}}}_{2}}={{\left( {\textbf{I}_{{{N}_{T}}D}}+\tilde{\lambda} \tilde{\boldsymbol{\Lambda} } \right)}^{-1}}\left( \tilde{\textbf{t}}+\tilde{\lambda}\big( {{{\tilde{\boldsymbol{\Gamma}}}}_{12}}{\left( {\bar{\textbf{v}}} \right)}\big)^{H} \right),\label{new_v2}
\end{equation}
where $\tilde{\lambda} \ge 0$ denotes the Lagrange multiplier. If $\tilde{\textbf{t}}$ satisfies the constraint (\ref{new_v2_constraint}), we have $\tilde{\lambda} = 0$ and $\tilde{\textbf{t}}$ is the optimal solution.  Otherwise, we have $\tilde{\lambda} > 0$ and the constraint (\ref{new_v2_constraint}) is satisfied with equality at the optimality of problem (\ref{new_v2_problem}). To find the optimal $\tilde{\lambda}$, we substitute (\ref{new_v2}) into the equality constraint $\tilde{\textbf{v}}_{2}^{H}\tilde{\boldsymbol{\Lambda} }{{\tilde{\textbf{v}}}_{2}}-2\mathcal{R}\left( {{{\tilde{\boldsymbol{\Gamma} }}}_{12}}\left( {\bar{\textbf{v}}} \right){{{\tilde{\textbf{v}}}}_{2}} \right)= -\overline{\rm MI}\left( {\bar{\textbf{v}}} \right)$ and obtain the following equation
\begin{equation}
\begin{aligned}
f\big( \tilde{\lambda}  \big)=&\sum\limits_{k=1}^{D{{N}_{T}}}{{\tilde{\mu }_{k}}{{\left| \frac{\tilde{t}_k+\tilde{\lambda} \tilde{\kappa}_{12}^{k}}{1+\tilde{\lambda} {\tilde{\mu }_{k}}} \right|}^{2}}}\\
&-2\mathcal{R}\left\{ \sum\limits_{k=1}^{D{{N}_{T}}}{{{\left( \tilde{\kappa }_{12}^{k} \right)}^{*}}\frac{\tilde{t}_k+\tilde{\lambda} \tilde{\kappa }_{12}^{k}}{1+\tilde{\lambda} {\tilde{\mu }_{k}}}} \right\}+\overline{\rm MI}\left( {\bar{\textbf{v}}} \right)=0,
\end{aligned}
\end{equation}
where $\tilde{t}_k$ and $\tilde{\kappa}_{12}^{k}$ are the $k$-th element of vectors 
$\tilde{\textbf{t}}$ and $\big({{{\tilde{\boldsymbol{\Gamma}}}}_{12}}{\left( {\bar{\textbf{v}}} \right)}\big)^{H}$, respectively; and $\left\{\tilde{\mu}_{k}\right\}$ denote the eigenvalues of  ${{\bar{\boldsymbol{\Gamma}}}_{22}}\left( {\bar{\textbf{v}}}\right)$. Since $f\big( \tilde{\lambda}  \big)$ decreases monotonically, the solution of $f\big( \tilde{\lambda}  \big)=0$ is unique. Then, the optimal $\tilde{\lambda}$ can be obtained using the bisection search. After finding $\tilde{\lambda}$, the optimal ${\rm{\textbf{v}}_{2}}$ can be calculated as
\begin{equation}\label{v2} 
{\rm{\textbf{v}}_{2}}=\textbf{Q}{{\tilde{\textbf{v}}}_{2}}.
\end{equation}

\emph{3)} ${\rm{\textbf{v}}}$ \emph{update:} The optimization problem w.r.t. ${\rm{\textbf{v}}}$ can be expressed as
\begin{equation}
\underset{\textbf{v}}{\mathop{\min}}\text{ } {\textbf{v}^{H}}\boldsymbol{\Pi} \textbf{v}+\frac{{\bar{\rho }}}{2}\sum\limits_{i=1}^{2}{\left\| {\textbf{v}_{i}}-\textbf{v}+{\textbf{c}_{i}} \right\|^{2}}.\label{v_update}
\end{equation}
Setting the gradient of the objective function in (\ref{v_update}) to zero, the optimal ${\rm{\textbf{v}}}$ is given by
\begin{equation}\label{v}
\textbf{v}={{\left( {\boldsymbol{\Pi} }+\bar{\rho }{\textbf{I}_{D{{N}_{T}}}} \right)}^{-1}}\left( \frac{{\bar{\rho }}}{2}\sum\limits_{i=1}^{2}{\left( {\textbf{v}_{i}}+{\textbf{c}_{i}} \right)} \right).
\end{equation}

Finally, the overall algorithm for solving $\mathcal{P}\text{1}$ is summarized in Algorithm 1.
\begin{figure}[!t]
	\removelatexerror
	\begin{algorithm}[H]
		\caption{Communication Precoder Design. } 
		{\bf Initialization:} {Initialize $\bar{\textbf{v}}={\rm vec}\left( {\bar{\textbf{V}}} \right)$ with ${\bar{\textbf{V}}}$ being the result of the previous outer iteration. Set the parameter $\bar{\rho }$ and calculate the inverse of the matrix $ {\boldsymbol{\Pi} }+\bar{\rho }{\textbf{I}_{D{{N}_{T}}}}$.}\\
		{\bf Repeat} [SCA Step]
		
		\hspace{0.6cm}Update ${\bar{\boldsymbol{\Gamma}}_{22}\left( {\bar{\textbf{v}}} \right)}$, ${\bar{\boldsymbol{\Gamma}}}_{12}\left( {\bar{\textbf{v}}} \right)$ and $\overline{\rm MI}\left( {\bar{\textbf{v}}} \right)$.
		
		\hspace{0.6cm}Perform eigen-decomposition on ${{\bar{\boldsymbol{\Gamma}}}_{22}}\left( {\bar{\textbf{v}}}\right)$.
		
		\hspace{0.6cm}Initialize ${\textbf{c}_{i}}\leftarrow \textbf{0},\forall i$ and $\textbf{v}=\bar{\textbf{v}}$. 
		
		\hspace{0.6cm}{\bf Repeat} [ADMM Step]
		
		\hspace{1.2cm}Update ${\textbf{v}_{1}}$ and ${{{\textbf{v}}}_{2}}$ according to (\ref{v1}) and (\ref{v2}).
		
		\hspace{1.2cm}Update $\textbf{v}$ according to (\ref{v}).
		
		\hspace{1.2cm}Update the dual variables ${\textbf{c}_{i}}\leftarrow {\textbf{c}_{i}}+ {\textbf{v}_{i}}-\textbf{v}, i=1,2$.
		
		\hspace{0.6cm}{\bf Until} ADMM convergence criterion is met.
		
		\hspace{0.6cm}Update $\bar{\textbf{v}}=\textbf{v}$.
		
		{\bf Until} SCA convergence criterion is met.
	\end{algorithm}
\end{figure}

Algorithm 1 is guaranteed to converge to a finite value of $\mathcal{P}\text{1}$. Firstly, the ADMM to the convex problem $\mathcal{P}\text{1.1}\big( \bar{\textbf{v}}\big)$ is convergent \cite{boyd2011distributed}. Secondly, the iterative optimization of $\mathcal{P}\text{1.1}\big( \bar{\textbf{v}}\big)$ in the SCA procedure is non-increasing and the objective function $r\left( \textbf{V} \right)$ is lower bounded by 0.

\emph{\textbf{Remark 1} (Complexity Analysis for Algorithm 1): }The first computation complexity comes from performing the eigen-decomposition on ${{\bar{\boldsymbol{\Gamma}}}_{22}}\left( {\bar{\textbf{v}}}\right)$ with a complexity of $\mathcal{O}( {{\left( D{{N}_{T}} \right)}^{3}})$, which can be done before the ADMM. The second computation complexity comes from performing the matrix inverse on $ {\boldsymbol{\Pi} }+\bar{\rho }{\textbf{I}_{D{{N}_{T}}}}$ with a complexity of $\mathcal{O}( {{\left( D{{N}_{T}} \right)}^{3}})$, which can be done before the SCA procedure. The third computation complexity comes from performing the matrix-vector multiplication with a complexity of $\mathcal{O}( {{\left( D{{N}_{T}} \right)}^{2}})$ when updating ${\rm{\textbf{v}}_{2}}$ and ${\rm{\textbf{v}}}$. Denote $N_{a1}$ as the number of iteration for the ADMM. Then, the total computational complexity of Algorithm 1 is $\mathcal{O}\big( ({{N}_{s1}+1}){{\left( D{{N}_{T}} \right)}^{3}}+2{{N}_{s1}}{{N}_{a1}}{{\left( D{{N}_{T}} \right)}^{2}} \big)$, which is much less than that of the IPM.

\section{Radar Waveform Design}
In this section, we optimize the radar waveform $\textbf{s}$ with fixed communication precoder $\textbf{V}$, that is,
\begin{equation}
\begin{aligned}
\mathcal{P}\text{2: }&\underset{\textbf{s}}{\mathop{\max}}\text{ }\sum\limits_{j=0}^{J}{\sigma _{\alpha ,j}^{2}}\frac{{\textbf{s}^{H}}\textbf{H}_{j}^{H}\textbf{w}{\textbf{w}^{H}}{\textbf{H}_{j}}\textbf{s}}{{\textbf{s}^{H}}\tilde{\textbf{R}}\textbf{s}+{r}\left( \textbf{V} \right)}\notag \\ 
&\text{s.t.}\quad(\rm \ref{com_MI}),(\rm \ref{radar_power}).
\end{aligned}
\end{equation}
$\mathcal{P}\text{2}$ is hard to tackle due to the non-convex constraint (\ref{com_MI}) and non-concave fractional objective function. Considering the application scene and the hardware limitation, some common waveform constraints (e.g., the similarity and PAPR constraints) should also be incorporated into waveform design. Different constraints will introduce different challenges. Since the waveform design previously developed for single path detection is not applicable to solving $\mathcal{P}\text{2}$ \cite{8352726,7202844}, we will develop effective waveform design algorithms using the SCA and SDP methods in the following.

\subsection{Waveform Design under Similarity Constraint}
The similarity constraint uses a known waveform $\textbf{s}_0$ as a benchmark and forces $\textbf{s}$ to share some good properties of $\textbf{s}_0$ in terms of the side-lobe levels and the envelope constancy. It can be written as \cite{7414411}
\begin{equation}
	\label{similarity}{{\left\| \textbf{s}-\varsigma {\textbf{s}_{0}} \right\|}^{2}}\le \epsilon {{P}_{R}},\text{ }{{\left| \epsilon  \right|}^{2}}\le 1,
\end{equation}
where $\textbf{s}_0$ satisfies ${{\left\| {\textbf{s}_{0}} \right\|}^{2}}={{P}_{R}}$, $\epsilon $ determines the level of the similarity and $\varsigma $ can be used to modulate the power of $\textbf{s}_0$. (\ref{similarity}) is equivalent to
\begin{equation}
	{\textbf{s}^{H}}\left( {\textbf{I}}_{KM_T}-\frac{{\textbf{s}_{0}}\textbf{s}_{0}^{H}}{{{P}_{R}}} \right)\textbf{s}\le \epsilon {{P}_{R}},\label{similarity_conversion}
\end{equation}
which is a convex quadratic constraint. 

To tackle the non-convexity of constraint (\ref{com_MI}), we introduce the following lemma.

\emph{\textbf{Lemma 4} (SCA-Based Transformation of Constraint} (\rm \ref{com_MI})\emph{): }(\rm \ref{com_MI}) is non-convex w.r.t. $\textbf{s}$ and can be successively approximated by the following convex quadratic constraint
\begin{equation}
{\textbf{s}^{H}}\hat{\boldsymbol{\Gamma} }\left( {\bar{\textbf{S}}} \right)\textbf{s}\le \widehat{\rm MI}\left( {\bar{\textbf{S}}} \right),\label{radar_SCA}
\end{equation}
where ${\bar{\textbf{S}}}=\bar{\textbf{s}}\bar{\textbf{s}}^{H}$ with $\bar{\textbf{s}}$ being the solution in the previous SCA iteration.
\begin{IEEEproof}
Define $\tilde{\textbf{S}}=\textbf{s}{\textbf{s}^{H}}$, ${\textbf{R}_{v}}=\sum\nolimits_{l=1}^{L}{}\sigma _{\upsilon ,l}^{2}\textbf{G}_{l}\textbf{V}{\textbf{V}^{H}}{\textbf{G}_{l}^{H}}$ and $\textbf{R}_{c}^{k}\big( \tilde{\textbf{S}} \big)=
\sigma _{c}^{2}{\textbf{I}_{{{N}_{R}}}}+\sum\nolimits_{i=1}^{I}{\sigma _{\gamma ,i}^{2}\textbf{G}_{rc}^{i}\left( k \right)\tilde{\textbf{S}}{{\left( \textbf{G}_{rc}^{i}\left( k \right) \right)}^{H}}}$. The function ${\rm MI}\left( \textbf{s},\textbf{V} \right)$ can be rewritten as ${\rm MI}\big( {\tilde{\textbf{S}}} \big)=\frac{1}{{\tilde{K}}}\sum\nolimits_{k=1}^{{\tilde{K}}}{}{\Big( \log \left| {\textbf{R}_{v}}+\textbf{R}_{c}^{k}\big( {\tilde{\textbf{S}}} \big) \right|-\log \left| \textbf{R}_{c}^{k}\big( {\tilde{\textbf{S}}} \big) \right| \Big)}$. ${\rm MI}\big( {\tilde{\textbf{S}}} \big)$ is convex w.r.t. $\tilde{\textbf{S}}$ and its first-order condition can be given by
\begin{equation}
{\rm MI}\big( \tilde{\textbf{S}}\big)\ge {\rm MI}\left( \bar{\textbf{S}}\right)-{\rm tr}\left( \hat{\boldsymbol{\Gamma}}\left( {\bar{\textbf{S}}} \right)\big( \tilde{\textbf{S}}-\bar{\textbf{S}} \big) \right),
\end{equation}
where \cite{8352726}
\begin{equation}
\begin{aligned}
\hat{\boldsymbol{\Gamma} }\big( {\bar{\textbf{S}}} \big)=&-\left( \frac{\partial {\rm MI}\big( {\tilde{\textbf{S}}} \big)}{\partial \tilde{\textbf{S}}} \right)_{\tilde{\textbf{S}}=\bar{\textbf{S}}}^{T}=\frac{1}{{\tilde{K}}}\sum\limits_{k=1}^{{\tilde{K}}}{}\bigg(\sum\limits_{i=1}^{I}{}\sigma _{\gamma ,i}^{2}{{\left( \textbf{G}_{rc}^{i}\left( k \right) \right)}^{H}}\\
&\Big[ {{\left( \textbf{R}_{c}^{k}\left( {\bar{\textbf{S}}} \right) \right)}^{-1}}-{{\left( {\textbf{R}_{v}}+\textbf{R}_{c}^{k}\left( {\bar{\textbf{S}}} \right) \right)}^{-1}} \Big]\textbf{G}_{rc}^{i}\left( k \right) \bigg)
\end{aligned}
\end{equation}
is the gradient of ${\rm MI}\big( \tilde{\textbf{S}}\big)$ at $\bar{\textbf{S}}$. Thus, the constraint (\ref{com_MI}) can be transformed to
\begin{equation}
{\rm MI}\left( \bar{\textbf{S}}\right)-{\rm tr}\left( \hat{\boldsymbol{\Gamma}}\left( {\bar{\textbf{S}}} \right)\left( \tilde{\textbf{S}}-\bar{\textbf{S}} \right) \right)\ge {\rm MI}_0.\label{new_s_constraint}
\end{equation}
Define 
\begin{equation}
\widehat{\rm MI}\left( {\bar{\textbf{S}}} \right)={\rm MI}\left( {\bar{\textbf{S}}} \right)+{\rm tr}\left( \hat{\boldsymbol{\Gamma} }\left( {\bar{\textbf{S}}} \right)\bar{\textbf{S}} \right)-{{\rm MI}_{0}}. 
\end{equation}
The constraint (\ref{new_s_constraint}) is equivalent to (\ref{radar_SCA}), which is convex due to the positive semidefinite matrix $\hat{\boldsymbol{\Gamma}}\left( {\bar{\textbf{S}}} \right)$. 	
\end{IEEEproof}

Next, we will deal with the non-concavity of the objective function. Note that the Dinkelbach-Type method in \cite{8352726,7202844} can be used for the fractional programming. However, it was only developed for single-ratio fractional program with quasi-concave objective functions and cannot apply for $\mathcal{P}\text{2}$ due to the lack of quasi-concavity of the sum-form fractional programming. Moreover, tedious iterations will reduce the efficiency of the algorithm. Thus, we develop a non-iterative SDP-based method to design radar waveform at a lower cost. Specifically, using Lemma 4, $\mathcal{P}\text{\rm 2}$ with similarity constraint (\ref{similarity_conversion}) can be approximated by a series of fractional SDP problems, i.e.,
\begin{subequations}
\begin{align}
\mathcal{P}\text{2.1}\big( \bar{\textbf{S}}\big)\text{: }&\underset{{\tilde{\textbf{S}}}}{\mathop{\max}}\text{ }\frac{{\rm tr}\big( \tilde{\boldsymbol{\Psi}}\tilde{\textbf{S}} \big)}{{\rm tr}\big(\tilde{\textbf{R}}\tilde{\textbf{S}} \big)+{{r}\left( \textbf{V} \right)}} \notag\\ 
&\text{s.t.}\quad{\rm tr}\big(\tilde{\textbf{S}} \big)\le {P}_{R}\label{power_constraint}\\ 
&\qquad\text{ }{\rm tr}\big( {\hat{\boldsymbol{\Gamma} }\left( {\bar{\textbf{S}}} \right)}\tilde{\textbf{S}} \big)\le \widehat{\rm MI}\left( {\bar{\textbf{S}}} \right)\label{com_constraint}\\
&\qquad\text{ } {\rm tr}\left(\left( {\textbf{I}}_{KM_T}-\frac{{\textbf{s}_{0}}\textbf{s}_{0}^{H}}{{{P}_{R}}} \right)\tilde{\textbf{S}}\right)\le \epsilon {{P}_{R}}\label{simi_constraint}\\
&\qquad\text{ } {\rm rank}(\tilde{\textbf{S}})=1\label{rank1_constraint}\\
&\qquad\text{ } \tilde{\textbf{S}}\succcurlyeq \textbf{0},\label{sd_constraint}
\end{align}
\end{subequations}
where $\tilde{\boldsymbol{\Psi}}=\sum\nolimits_{j=0}^{J}{\sigma _{\alpha ,j}^{2} \textbf{H}_{j}^{H}\textbf{w}{\textbf{w}^{H}}{\textbf{H}_{j}}}$.

After relaxing the rank-one constraint (\ref{rank1_constraint}), $\mathcal{P}\text{2.1}\big( \bar{\textbf{S}}\big)$ can be solved with the Charnes-Cooper transformation \cite{charnes1962programming}. However, this method can not guarantee an optimal solution to $\mathcal{P}\text{2.1}\big( \bar{\textbf{S}}\big)$. To optimally solve it, we construct the following dual problem
\begin{equation}
\begin{aligned}
\mathcal{P}\text{2.2}\big( \bar{\textbf{S}}\big)\text{: }&\underset{{\tilde{\textbf{S}}}}{\mathop{\min}}\text{ }{\rm tr}\big( \tilde{\textbf{S}} \big)\\ 
&\text{s.t.}\quad\frac{{\rm tr}\big( \tilde{\boldsymbol{\Psi} }\tilde{\textbf{S}} \big)}{{\rm tr}\big( \tilde{\textbf{R}}\tilde{\textbf{S}} \big)+{{{r}\left( \textbf{V} \right)}}}\ge \tilde{p} \\ 
&\qquad\text{ }(\rm \ref{com_constraint}), (\ref{simi_constraint}), (\rm \ref{sd_constraint})
\end{aligned}
\end{equation}
and explore the relationship between the two problems, where $\tilde{p}$ is a non-negative constant. For the SDP problems $\mathcal{P}\text{2.1}\big( \bar{\textbf{S}}\big)$ and $\mathcal{P}\text{2.2}\big( \bar{\textbf{S}}\big)$, we have the following result.

\emph{\textbf{Proposition 2} (Optimal Solution of} $\mathcal{P}\text{2.2}\big( \bar{\textbf{S}}\big)$ \emph{under Multi-path Detection}\emph{): }If $\tilde{p}$ equals to the optimal value of $\mathcal{P}\text{2.1}\big( \bar{\textbf{S}}\big)$ after relaxing (\rm\ref{rank1_constraint}), the rank-one optimal solution of $\mathcal{P}\text{2.2}\big( \bar{\textbf{S}}\big)$ exists and is also optimal to $\mathcal{P}\text{2.1}\big( \bar{\textbf{S}}\big)$.
\begin{IEEEproof}
	After relaxing (\rm\ref{rank1_constraint}), $\mathcal{P}\text{2.1}\big( \bar{\textbf{S}}\big)$ can be converted to
	\begin{equation}
	\begin{aligned}
		\mathcal{P}\text{2.3}\big( \bar{\textbf{S}}\big)\text{: }&\underset{{\tilde{\textbf{S}}}}{\mathop{\max}}\text{ }\frac{{\rm tr}\big( \tilde{\boldsymbol{\Psi}}\tilde{\textbf{S}} \big)}{{\rm tr}\big(\tilde{\textbf{R}}\tilde{\textbf{S}} \big)+{{r}\left( \textbf{V} \right)}} \\ 
		&\text{s.t.}(\rm\ref{power_constraint}), (\rm \ref{com_constraint}), (\ref{simi_constraint}), (\rm \ref{sd_constraint}).
	\end{aligned}
	\end{equation}
	Let $\tilde{p}$ and ${{\tilde{\textbf{S}}}_{1}}$ be the optimal value and optimal solution to $\mathcal{P}\text{2.3}\big( \bar{\textbf{S}}\big)$, respectively. We further solve $\mathcal{P}\text{2.2}\big( \bar{\textbf{S}}\big)$ to obtain its optimal value $\tilde{q}$ and optimal solution ${{\tilde{\textbf{S}}}_{2}}$.
		
	We first prove that $\mathcal{P}\text{2.2}\big( \bar{\textbf{S}}\big)$ and $\mathcal{P}\text{2.3}\big( \bar{\textbf{S}}\big)$ share the same optimal solutions. Specifically, it is easy to verify that ${{\tilde{\textbf{S}}}_{1}}$ is a feasible solution to $\mathcal{P}\text{2.2}\big( \bar{\textbf{S}}\big)$. Then, we have ${\rm tr}\big( {{{\tilde{\textbf{S}}}}_{2}} \big)\le {\rm tr}\big( {{{\tilde{\textbf{S}}}}_{1}} \big)\le {P}_{R}$. Thus, ${{\tilde{\textbf{S}}}_{2}}$ is a feasible solution to $\mathcal{P}\text{2.3}\big( \bar{\textbf{S}}\big)$, which implies that ${{\rm tr}\big( \tilde{\boldsymbol{\Psi} }{{{\tilde{\textbf{S}}}}_{2}} \big)}/{\big( {\rm tr}\big( \tilde{\textbf{R}}{{{\tilde{\textbf{S}}}}_{2}} \big)+{{{r}\left( \textbf{V} \right)}} \big)}\le \tilde{p}$. Since ${{\tilde{\textbf{S}}}_{2}}$ is the optimal solution to $\mathcal{P}\text{2.2}\big( \bar{\textbf{S}}\big)$, we have ${{\rm tr}\big( \tilde{\boldsymbol{\Psi} }{{{\tilde{\textbf{S}}}}_{2}} \big)}/{\big( {\rm tr}\big( \tilde{\textbf{R}}{{{\tilde{\textbf{S}}}}_{2}} \big)+{{{r}\left( \textbf{V} \right)}} \big)}\ge \tilde{p}$. Hence, ${{\rm tr}\big( \tilde{\boldsymbol{\Psi} }{{{\tilde{\textbf{S}}}}_{2}} \big)}/{\big( {\rm tr}\big( \tilde{\textbf{R}}{{{\tilde{\textbf{S}}}}_{2}} \big)+{{r}\left( \textbf{V} \right)} \big)}=\tilde{p}$ holds, i.e., the optimal solution to $\mathcal{P}\text{2.2}\big( \bar{\textbf{S}}\big)$ is also optimal to $\mathcal{P}\text{2.3}\big( \bar{\textbf{S}}\big)$.	
	
	Using the conclusion in \cite[Theorem 3.2]{huang2009rank}, we can conclude that $\mathcal{P}\text{2.2}\big( \bar{\textbf{S}}\big)$ has a optimal solution satisfying ${\rm rank}^{2}(\tilde{\textbf{S}})\le 3$, i.e., there exists a rank-one optimal solution to $\mathcal{P}\text{2.2}\big( \bar{\textbf{S}}\big)$, which is also optimal to $\mathcal{P}\text{2.3}\big( \bar{\textbf{S}}\big)$ and $\mathcal{P}\text{2.1}\big( \bar{\textbf{S}}\big)$	
\end{IEEEproof}

Notice that Proposition 2 only states the existence of a globally optimal rank-one solution to $\mathcal{P}\text{2.2}\big( \bar{\textbf{S}}\big)$. In general, the global optimal solution may not be unique, and the convex optimization solvers may not provide a rank-one solution. If ${\rm rank}(\tilde{\textbf{S}}_{2})\ge 2$, we can exploit the rank reduction procedures proposed in \cite{huang2009rank} to generate a rank-one optimal solution. Moreover, in a special case that the radar detects the target only along the LoS, i.e., $J=0$, the optimal solution of $\mathcal{P}\text{2.2}\big( \bar{\textbf{S}}\big)$ is always rank one, as shown in the following corollary.

\emph{\textbf{Corollary 1} (Optimal Solution of }$\mathcal{P}\text{2.2}\big( \bar{\textbf{S}}\big)$ \emph{under LoS Detection}\emph{):} When $J=0$, the optimal solution of $\mathcal{P}\text{2.2}\big( \bar{\textbf{S}}\big)$ is always rank one.
\begin{IEEEproof}
See Appendix C.
\end{IEEEproof}

Finally, the overall algorithm is summarized in Algorithm 2.

We can guarantee the convergence of Algorithm 2. {Firstly, each convex approximation problem in the SCA procedure can be solved optimally, which guarantees the iterative optimization of $\mathcal{P}\text{2.1}\big( \bar{\textbf{S}}\big)$ is non-decreasing. Secondly, the optimal value of $\mathcal{P}\text{2}$ is upper-bounded, i.e.,}
$$\sum\limits_{j=0}^{J}{\sigma _{\alpha ,j}^{2}\frac{{\textbf{s}^{H}}\textbf{H}_{j}^{H}\textbf{w}{\textbf{w}^{H}}{\textbf{H}_{j}}\textbf{s}}{{\textbf{s}^{H}}\tilde{\textbf{R}}\textbf{s}+{r}\left( \textbf{V} \right)}}\le\frac{{\textbf{s}^{H}}\tilde{\boldsymbol{\Psi} }\textbf{s}}{{r}\left( \textbf{V} \right)}\le \hat{\mu}\frac{{P}_{R}}{{r}\left( \textbf{V} \right)},$$
where $\hat{\mu}$ denotes the maximum eigenvalue of $\tilde{\boldsymbol{\Psi} }$. 
Thus, the proposed algorithms can converge to a finite value.

\emph{\textbf{Remark 2} (Complexity Analysis for Algorithm 2): }
The computational overhead mainly comes from solving the SDP problems $\mathcal{P}\text{2.2}\big( \bar{\textbf{S}}\big)$ and $\mathcal{P}\text{2.3}\big( \bar{\textbf{S}}\big)$. Exploiting the IPM to solve each SDP problem requires $\mathcal{O}\left( \sqrt{3}\log \left( {1}/{\varepsilon } \right) \right)$ iterations to converge, where ${\varepsilon}$ denotes the relative accuracy and each iteration has a computational complexity of $\mathcal{O}\left( M_{T}^{3.5}{{K}^{3.5}} \right)$ \cite{nesterov1994interior}. Denote $N_{s2}$ as the number of iteration of the SCA method in Algorithm 2. Then, the total computational complexity is $\mathcal{O}\left( 2\sqrt{3}{{N}_{s2}}M_{T}^{3.5}{{K}^{3.5}}\log \left( {1}/{\varepsilon } \right) \right)$. 

\begin{figure}[!t]
	\removelatexerror
	\begin{algorithm}[H]
		\caption{Waveform Design under Similarity Constraint.} 
		{\bf Initialization:} {Initialize ${\bar{\textbf{S}}}=\bar{\textbf{s}}\bar{\textbf{s}}^{H}$ with $\bar{\textbf{s}}$ being the solution of the previous outer iteration.}\\
		{\bf Repeat} [SCA]
		
		\hspace{0.6cm}Solve $\mathcal{P}\text{2.3}\big( \bar{\textbf{S}}\big)$ to find its optimal value $\tilde{p}$.
		
		\hspace{0.6cm}Solve $\mathcal{P}\text{2.2}\big( \bar{\textbf{S}}\big)$ to find its optimal solution ${\tilde{\textbf{S}}}_{2}$.
		
		\hspace{0.6cm}Evaluate $\tilde{R}={\rm rank}(\tilde{\textbf{S}}_{2})$.
		
		\hspace{0.6cm}{\bf While} $\tilde{R}>1$ {\bf do} [Rank Reduction]
		
		\hspace{1.2cm}Decompose ${{\tilde{\textbf{S}}}_{2}}={{\tilde{\textbf{U}}}_{2}}\tilde{\textbf{U}}_{2}^{H}$, ${{\tilde{\textbf{U}}}_{2}}\in {{\mathbb{C}}^{K{{M}_{T}}\times \tilde{R}}}$.
		
		\hspace{1.2cm}Find a nonzero solution ${\boldsymbol{\Lambda}}$ of linear equations
		\begin{equation}\nonumber		
			\left\{ \begin{matrix}\begin{aligned}
					&{\rm tr}\left( \tilde{\textbf{U}}_{2}^{H}\left( \tilde{p}\tilde{\textbf{R}}-{\boldsymbol{\Psi}}  \right){{{\tilde{\textbf{U}}}}_{2}}{\boldsymbol{\Lambda}}  \right)=0  \\
					&{\rm tr}\left( \tilde{\textbf{U}}_{2}^{H}{{{\tilde{\textbf{U}}}}_{2}}{\boldsymbol{\Lambda}}  \right)=0\\
					&{\rm tr}\left(
					\tilde{\textbf{U}}_{2}^{H}\left( {\textbf{I}}_{KM_T}-\frac{{\textbf{s}_{0}}\textbf{s}_{0}^{H}}{{{P}_{R}}}  \right){{{\tilde{\textbf{U}}}}_{2}}{\boldsymbol{\Lambda}}
					\right)=0		
	  \end{aligned}
			\end{matrix} \right.,
		\end{equation}
		\hspace{1.2cm}where ${\boldsymbol{\Lambda}}$ is a $\tilde{R}\times\tilde{R}$ Hermitian matrix.
		
		\hspace{1.2cm}Evaluate the eigenvalues ${{\tilde{\delta }}_{1}},\ldots,{{\tilde{\delta }}_{{\tilde{R}}}}$ of ${\boldsymbol{\Lambda}}$.
		
		\hspace{1.2cm}Determine $\big| {{{\tilde{\delta }}}_{{{{\tilde{r}}}_{0}}}} \big|=\max \big\{ \big| {{{\tilde{\delta }}}_{{\tilde{r}}}} \big|:1\le \tilde{r}\le \tilde{R} \big\}$.
		
		\hspace{1.2cm}Compute ${{\tilde{\textbf{S}}}_{2}}={{\tilde{\textbf{U}}}_{2}}\big( {\textbf{I}_{{\tilde{R}}}}-\big( 1/\big| {{{\tilde{\delta }}}_{{{{\tilde{r}}}_{0}}}} \big| \big){\boldsymbol{\Lambda}} \big)\tilde{\textbf{U}}_{2}^{H}$
		
		\hspace{1.2cm}Evaluate $\tilde{R}={\rm rank}(\tilde{\textbf{S}}_{2})$.
		
		\hspace{0.6cm}{\bf End while}
		
		\hspace{0.6cm}Update $\bar{\textbf{S}}=\tilde{\textbf{S}}_{2}$.	
		
		{\bf Until} SCA convergence criterion is met.
		
		Perform eigen-decomposition on ${\bar{\textbf{S}}}$ as ${{ \bar{\mathrm{\textbf{S}}}}}={{\textbf{s}}}{{\textbf{s}}^{H}}$.
	\end{algorithm}
\end{figure}

\subsection{Waveform Design under PAPR Constraints} 
High PAPR values require linear amplifiers with large dynamic range, which is unrealistic in modern radar systems. Thus, we bound the PAPR value at each antenna by the following constraints \cite{wu2017transmit}, \cite{5732713}.
\begin{subequations}\label{PAPR}
\begin{align}
&\label{PAPR1}{{\left\| \textbf{s} \right\|}^{2}}={{P}_{R}}  \\
&\label{PAPR2}\underset{1\le n\le K{{M}_{T}}}{\mathop{\max }}\left\{ {{\left| {{s}_{n}} \right|}^{2}} \right\}\le \frac{\eta {{P}_{R}}}{K{{M}_{T}}},
\end{align}
\end{subequations}
where $\eta$ controls the maximum allowable PAPR value and ${{s}_{n}}$ denotes the \emph{n}-th element of the waveform $\textbf{s}$. In particular, when $\eta=1$, the PAPR constraint reduces to the constant modulus constraint. Define ${\textbf{F}_{n}}\in{{\mathbb{C}}^{K{{M}_{T}}\times K{{M}_{T}}}}$ with the $\left( n,n \right)$-th element being ${{KM}_{T}}/\left( \eta {{P}_{R}} \right)$ and other elements being 0. Then, the constraint (\ref{PAPR2}) can be rewritten as
\begin{equation}\label{PAPR3}
{\textbf{s}^{H}}{{{{\textbf{F}}}}_{n}}\textbf{s}\le 1,\forall n=1,\ldots ,K{{M}_{T}}.
\end{equation}

The waveform design problem is thus given by
\begin{equation}
\begin{aligned}
\mathcal{P}\text{3: }&\underset{\textbf{s}}{\mathop{\max}}\text{ }\sum\limits_{j=0}^{J}{\sigma _{\alpha ,j}^{2}}\frac{{\textbf{s}^{H}}\textbf{H}_{j}^{H}\textbf{w}{\textbf{w}^{H}}{\textbf{H}_{j}}\textbf{s}}{{\textbf{s}^{H}}\tilde{\textbf{R}}\textbf{s}+{r}\left( \textbf{V} \right)}\notag \\ 
&\text{s.t.}\quad(\rm \ref{com_MI}),(\rm \ref{PAPR1}),(\ref{PAPR3}),
\end{aligned}
\end{equation}
which is hard to tackle due to the non-concave objective function and non-convex constraints (\rm \ref{com_MI}) and (\rm \ref{PAPR1}). We can solve $\mathcal{P}{3}$ by the same SCA procedure in Algorithm 2. During each iteration, we need to solve the following fractional SDP problem
\begin{subequations}
\begin{align}
\mathcal{P}\text{3.1}\big( \bar{\textbf{S}}\big)\text{: }&\label{papr_obj}\underset{{\tilde{\textbf{S}}}}{\mathop{\max }}\text{ }\frac{\text{\rm tr}\big( \tilde{\boldsymbol{\Psi} }\tilde{\textbf{S}} \big)}{{\rm tr}\big( \breve{\textbf{R}}\tilde{\textbf{S}} \big)} \\ 
&\label{papr_MI}\text{s.t.}\quad{\rm tr}\big( \tilde{\boldsymbol{\Gamma} }\left( {\bar{\textbf{S}}} \right)\tilde{\textbf{S}} \big)\le 1 \\ 
&\label{papr_po}\qquad\text{ }{\rm tr}\big( {\tilde{\textbf{S}}} \big)={{P}_{R}} \\ 
&\label{papr_F}\qquad\text{ }{\rm tr}\big( {{{\tilde{\textbf{F}}}}_{n}}\tilde{\textbf{S}} \big)\le 1, \forall n\\ 
&\qquad\text{ }(\rm\ref{rank1_constraint}), (\rm\ref{sd_constraint}),\notag
\end{align}
\end{subequations}
where (\ref{papr_obj}) comes from the original objective function and (\rm\ref{PAPR1}) with $\breve{\textbf{R}}=\tilde{\textbf{R}}+\left( {{r}\left( \textbf{V} \right)}/{{{P}_{R}}} \right)\textbf{I}_{KM_T}$; (\ref{papr_MI}) comes from (\ref{radar_SCA}) and (\rm\ref{PAPR1}) with $\tilde{\boldsymbol{\Gamma} }\left( {\bar{\textbf{S}}} \right)={\big( \hat{\boldsymbol{\Gamma} }\left( {\bar{\textbf{S}}} \right)+{\textbf{I}_{K{{M}_{T}}}} \big)}/\big({\widehat{\rm MI}\left( {\bar{\textbf{S}}} \right)+{{P}_{R}}}\big)$; and (\ref{papr_F}) comes from (\ref{PAPR3}) and (\rm\ref{PAPR1}) with ${{\tilde{\textbf{F}}}_{n}}={\big( {\textbf{F}_{n}}+{\textbf{I}_{K{{M}_{T}}}} \big)}/\big({1+{{P}_{R}}}\big), \forall n$. It is easy to verify that $\breve{\textbf{R}}$, $\tilde{\boldsymbol{\Gamma} }\left( {\bar{\textbf{S}}} \right)$ and ${\{{\tilde{\textbf{F}}}_{n}\}}_{n=1}^{KM_T}$ are all positive definite. 

After relaxing the rank-one constraint (\rm\ref{rank1_constraint}), $\mathcal{P}\text{3.1}\big( \bar{\textbf{S}}\big)$ can be solved by exploiting the Charnes-Cooper transformation. However, due to the large number of linear constraints on $\tilde{\textbf{S}}$, the convex relaxation of $\mathcal{P}\text{3.1}\big( \bar{\textbf{S}}\big)$ does not guarantee the existence of a rank-one optimal solution. Thus, eigen-decomposition and randomization are required to obtain a rank-one sub-optimal solution. Since randomization based methods can sometimes lead to prohibitively high computational complexity, we provide a more efficient design.

We first consider $\mathcal{P}\text{3.1}\big( \bar{\textbf{S}}\big)$ without constraints (\rm\ref{papr_F}), i.e.,
\begin{equation}
	\begin{aligned}
		\mathcal{P}\text{3.2}\big( \bar{\textbf{S}}\big)\text{: }&\underset{{\tilde{\textbf{S}}}}{\mathop{\max }}\text{ }\frac{\text{\rm tr}\big( \tilde{\boldsymbol{\Psi} }\tilde{\textbf{S}} \big)}{{\rm tr}\big( \breve{\textbf{R}}\tilde{\textbf{S}} \big)} \\
		&\text{s.t.}\quad(\rm\ref{papr_MI}), (\rm\ref{papr_po}), (\rm\ref{rank1_constraint}), (\rm\ref{sd_constraint}).
	\end{aligned}
\end{equation}
After Charnes-Cooper transformation and the relaxation of rank-one constraint (\rm\ref{rank1_constraint}), $\mathcal{P}\text{3.2}\big( \bar{\textbf{S}}\big)$ can be converted to
\begin{equation}
	\begin{aligned}
		\mathcal{Q}\text{3.2}\big( \bar{\textbf{S}}\big)\text{: }&\underset{{\hat{\textbf{T}},\hat{t}}}{\mathop{\max }}\text{ }\text{\rm tr}\big( \tilde{\boldsymbol{\Psi} }\hat{\textbf{T}} \big) \\ 
		&\text{s.t.}\quad{\rm tr}\big( \tilde{\boldsymbol{\Gamma} }\left( {\bar{\textbf{S}}} \right)\hat{\textbf{T}} \big)\le \hat{t} \\ 
		&\qquad\text{ }{{\rm tr}\big( \breve{\textbf{R}}\hat{\textbf{T}} \big)}\le1\\
		&\qquad\text{ }{\rm tr}\big( {\hat{\textbf{T}}} \big)=\hat{t}{{P}_{R}} \\ 
		&\qquad\text{ } \hat{\textbf{T}}\succcurlyeq \textbf{0},\hat{t}\ge 0.
	\end{aligned}
\end{equation}
For the problems $\mathcal{P}\text{3.2}\big( \bar{\textbf{S}}\big)$ and $\mathcal{Q}\text{3.2}\big( \bar{\textbf{S}}\big)$, we have the following result.

\emph{\textbf{Proposition 3} (Optimal Solution of} $\mathcal{P}\text{3.2}\big( \bar{\textbf{S}}\big)$\emph{):}
There exits an optimal solution $(\hat{\textbf{T}}^{\star},\hat{t}^{\star})$ for $\mathcal{Q}\text{3.2}\big( \bar{\textbf{S}}\big)$ that makes $\hat{\textbf{T}}^{\star}/\hat{t}^{\star}$ optimal for $\mathcal{P}\text{3.2}\big( \bar{\textbf{S}}\big)$.
\begin{IEEEproof}
Similar to $\mathcal{P}\text{2.2}\big( \bar{\textbf{S}}\big)$, $\mathcal{Q}\text{3.2}\big( \bar{\textbf{S}}\big)$ also has an optimal solution $\hat{\textbf{T}}^{\star}$ of rank one. Perform rank reduction procedures can obtain the solution. Since Charnes-Cooper transformation is a equivalent conversion, $\hat{\textbf{T}}^{\star}/\hat{t}^{\star}$ is optimal for $\mathcal{P}\text{3.2}\big( \bar{\textbf{S}}\big)$.
\end{IEEEproof}

Upon obtaining the optimal solution ${{\tilde{\textbf{S}}}_{3}}={{\tilde{\textbf{s}}}_{3}}\tilde{\textbf{s}}_{3}^{H}$ of $\mathcal{P}\text{3.2}\big( \bar{\textbf{S}}\big)$, we need to project ${\tilde{\textbf{s}}}_{3}$ into PAPR constraints, i.e.,
\begin{equation}\label{project}\underset{\textbf{s}}{\mathop{\min }}\,\left\| \textbf{s}-{{{\tilde{\textbf{s}}}}_{3}} \right\|\text{ s.t. }(\rm\ref{PAPR1}), (\rm\ref{PAPR3}).\end{equation}
This is a matrix nearness problem and we can exploit the alternating projection method in \cite{1377501} to solve it. We omit it here for conciseness. 

In the special case of LoS detection, the optimal solution to $\mathcal{P}\text{3.1}\big( \bar{\textbf{S}}\big)$ can be find by the following corollary.

\emph{\textbf{Corollary 2} (Optimal Solution of }$\mathcal{P}\text{3.1}\big( \bar{\textbf{S}}\big)$ \emph{under LoS Detection}\emph{):} We can solve the rank-one semidefinite convex relaxation of $\mathcal{P}\text{3.1}\big( \bar{\textbf{S}}\big)$ to obtain its optimal solution.
\begin{IEEEproof}
The optimal solution to $\mathcal{P}\text{3.1}\big( \bar{\textbf{S}}\big)$ after relaxing the rank-one constraint can be calculated based on the Charnes-Cooper transformation. Applying the conclusion in \cite[Proposition 2]{8756034}, we can prove that the solution is rank-one.
\end{IEEEproof}
\begin{figure}[!t]
	\removelatexerror
	\begin{algorithm}[H]
		\caption{Waveform Design under PAPR Constraints.} 
		{\bf Initialization:} {Initialize ${\bar{\textbf{S}}}=\bar{\textbf{s}}\bar{\textbf{s}}^{H}$ with $\bar{\textbf{s}}$ being the solution of the previous outer iteration.}\\
		{\bf Repeat} [SCA]
		
		\hspace{0.6cm}Solve $\mathcal{Q}\text{3.2}\big( \bar{\textbf{S}}\big)$ to find its optimal solution $(\hat{\textbf{T}}^{\star},\hat{t}^{\star})$.
		
		\hspace{0.6cm}Decompose $\hat{\textbf{T}}^{\star}/\hat{t}^{\star}={{\tilde{\textbf{s}}}_{3}}\tilde{\textbf{s}}_{3}^{H}$.
		
		\hspace{0.6cm}Solve (\rm\ref{project}) to obtain $\bar{\textbf{s}}$ satisfying PAPR constraints.

		\hspace{0.6cm}Update $\bar{\textbf{S}}=\bar{\textbf{s}}\bar{\textbf{s}}^{H}$.	
		
		{\bf Until} SCA convergence criterion is met.
		
		Perform eigen-decomposition on ${\bar{\textbf{S}}}$ as ${{ \bar{\mathrm{\textbf{S}}}}}={{\textbf{s}}}{{\textbf{s}}^{H}}$.
	\end{algorithm}
\end{figure}

Finally, the overall algorithm is summarized in Algorithm 3. The convergence analysis is similar to that of Algorithm 2.

\emph{\textbf{Remark 3} (Complexity Analysis for Algorithm 3): }
The computational overhead mainly comes from solving the SDP problem $\mathcal{Q}\text{3.2}\big( \bar{\textbf{S}}\big)$. Denote $N_{s3}$ as the number of iteration of the SCA method in Algorithm 3. Then, the total computational complexity is $\mathcal{O}\left( \sqrt{3}{{N}_{s3}}M_{T}^{3.5}{{K}^{3.5}}\log \left( {1}/{\varepsilon } \right) \right)$.

\section{Numerical Results And Discussion}
In this section, we will provide numerical examples to evaluate the performances of the proposed algorithms. The number of antennas at the radar transmitter, the radar receiver, the BS and the CU are $M_T=8$, $M_R=18$, $N_T=10$ and $N_R=4$, respectively. The radar system has the PRI of $\tilde{K}=60$, the pulse duration of $K=4$ and maximum transmit power of $P_R=10$ walts. The communication system has the number of data streams of $D=4$ and maximum transmit power of $P_B=1$ walt.  The channel parameter settings in the simulations are provided as follows:
\begin{itemize}[topsep = 0 pt]
	\item The target is located in the angular direction of ${{\theta }_{0}}={{20}^{\circ}}$. There are $J=3$ scattering patches causing indirect path returns, which are located in the direction of ${{\theta }_{1}}={{-10}^{\circ}}$, ${{\theta }_{2}}={{-17}^{\circ}}$ and ${{\theta }_{3}}={{-25}^{\circ}}$, respectively. The signal-to-noise ratio along the direct path, given by ${{\rm SNR}_{r,d}}={P_R}{\sigma _{\alpha ,0}^{2}}/{\sigma _{r}^{2}}$, is set as 20dB, and that along indirect paths, denoted by ${{\rm SNR}_{r,id}}={P_R}{\sigma_{\alpha,j}^{2}}/{\sigma _{r}^{2}},\forall j$,  is set as 18dB.
	\item There are $Q=5$ scatterers causing clutter returns with $\tilde{\tau}_q$ and ${\tilde{\theta }_{q}}$ randomly chosen in $\big\{ -K+1,\ldots {K-1} \big\}$ and $({0}^{\circ},10^{\circ}) $, respectively. The clutter-to-noise ratio, denoted by ${{\rm CNR}_{r}}={P_R}{\sigma_{\tilde{\alpha},q}^{2}}/{\sigma _{r}^{2}},\forall q$, is set as 30dB.
	\item There are $G=6$ paths between the BS and the radar receiver with $\phi_{cr}^{g}$ and $\varphi _{cr}^{g}$ randomly chosen in $({30}^{\circ},50^{\circ}) $ and $({-90}^{\circ},90^{\circ}) $, respectively. The interference-to-noise ratio, denoted by ${{\rm INR}_{r}}={P_B}{\sigma _{\beta ,g}^{2}}/{\sigma _{r}^{2}},\forall g$, is set to 20dB.
	\item There are $L=3$ paths between the BS and the CU with $\vartheta _{r}^{l}$ and $\vartheta _{t}^{l}$ randomly chosen in $({-90}^{\circ},90^{\circ})$ and $({-90}^{\circ},90^{\circ})$, respectively. The signal-to-noise ratio at the CU, denoted by ${{\rm SNR}_{c}}={P_B}{\sigma _{\upsilon ,l}^{2}}/{\sigma _{c}^{2}},\forall l$, is set as 25dB.
	\item There are $I=6$ paths between the radar and the CU with ${\tau}_{rc}^{i}$, $\varphi_{rc}^{i}$ and $\phi_{rc}^{i}$ randomly chosen in $\big\{ 0,\ldots \tilde{K}-1 \big\}$, $({-10}^{\circ},20^{\circ})$ and $({30}^{\circ},70^{\circ})$, respectively. The interference-to-noise ratio at the CU, denoted by ${{\rm INR}_{c}}={P_R}{\sigma_{\gamma ,i}^{2} }/{\sigma _{c}^{2}},\forall i$, is set as 40dB.
\end{itemize}
In the alternating optimization algorithms, we set $\bar{\rho}=100$, ${\rm MI}_0=7$ [nats per symbol], $\epsilon=0.7$ and $\eta=3$. The orthogonal linear frequency modulation waveform is chosen as the initial radar waveform, denoted by $\textbf{s}_0$, whose space-time matrix is given by \cite{8352726,7762192}
\begin{equation}\notag
{\textbf{S}_{0}}\left( m,k \right)=\frac{\sqrt{P_R}{{e}^{{j2\pi m\left( k-1 \right)}/{{{M}_{T}}}}}{{e}^{{j\pi {{\left( k-1 \right)}^{2}}}/{{{M}_{T}}}}}}{\sqrt{{{M}_{T}}K}},
\end{equation}
where $m=1,\cdots ,{{M}_{T}}$ and $k=1,\cdots ,K$. Then, we can obtain ${\textbf{s}_{0}}={\rm vec}\left( {\textbf{S}_{0}} \right)$. The initial communication precoder is given by ${\textbf{V}_{0}}=\sqrt{{{{P}_{B}}}/{D}}{{\left[{\textbf{I}_{D}}\text{ } {\textbf{0}_{D\times \left( {{N}_{T}}-D \right)}} \right]}^{T}}$. Thus, the initial radar space-time filter can be calculated as $\textbf{w}_0=\mathcal{M}\left( {\textbf{R}^{-1}}\left( \textbf{V}_0,\textbf{s}_0 \right)\boldsymbol{\Psi} \left( \textbf{s}_0 \right) \right)$. The proposed algorithms terminate when the successive difference of the objective function values is less than $10^{-3}$.

\begin{figure}[!t]
	\centering
	\includegraphics[width=0.43\textwidth]{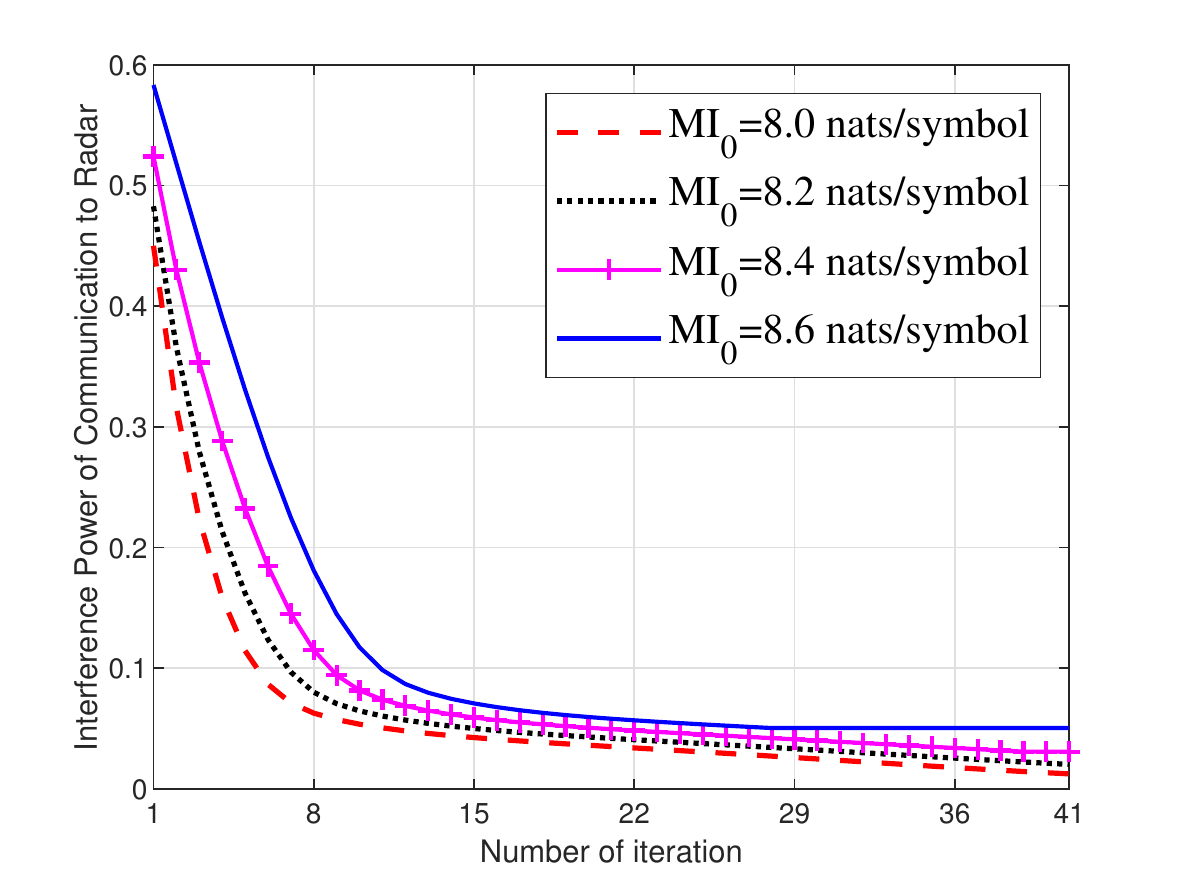}	
	\caption{{Objective function values in $\mathcal{P}\text{1}$ versus the number of iterations of Algorithm 1 when ${\rm MI_0}$ = 8.0, 8.2, 8.4 and 8.6 nats per symbol.}}\label{A1_converge}
\end{figure}

\begin{figure}[!t]
	\centering
	\includegraphics[width=0.43\textwidth]{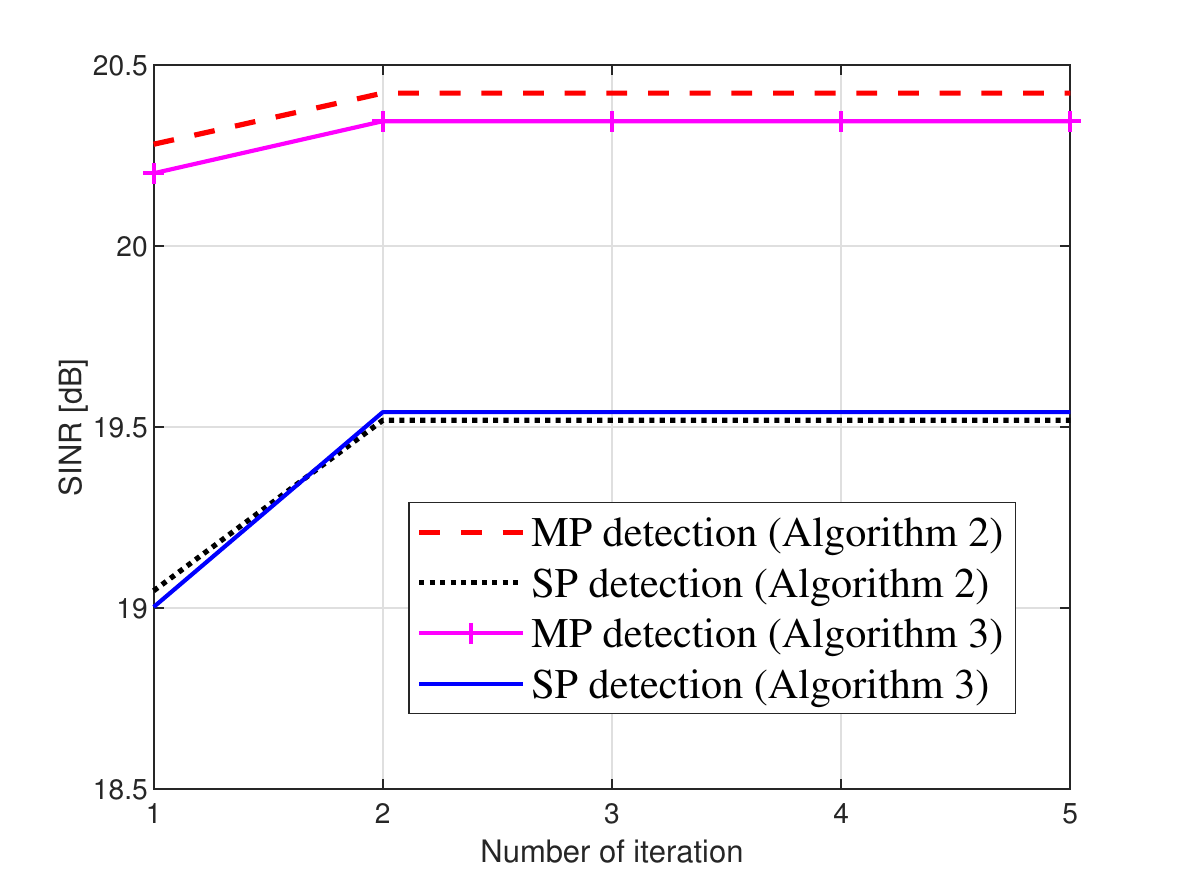}	
	\caption{{The radar output SINR versus the number of iterations of Algorithm 2 and Algorithm 3, respectively.}}\label{A23_converge}
\end{figure}

{We evaluate the convergence performance of the proposed
algorithms in Fig. \ref{A1_converge} and Fig. \ref{A23_converge}. It can be observed from Fig. \ref{A1_converge} that the objective function values in $\mathcal{P}\text{1}$ decrease continuously during the iteration until Algorithm 1 converges. Additionally, Fig. \ref{A1_converge} clearly shows that relaxing the communication rate constraint will result in a smaller objective function value. This is because a smaller ${\rm MI_0}$ will bring larger feasible sets for the optimization problem $\mathcal{P}\text{1}$. Fig. \ref{A23_converge} verifies the convergence of Algorithm 2 and Algorithm 3.} As expected, combining multi-path (MP for short) echoes yields higher SINR values than using single-path (SP for short) detection alone. This is because the existence of multi-path provides extra information on the target and thus increases the spatial diversity of the radar.

\begin{figure}[!t]
	\centering
	\includegraphics[width=0.43\textwidth]{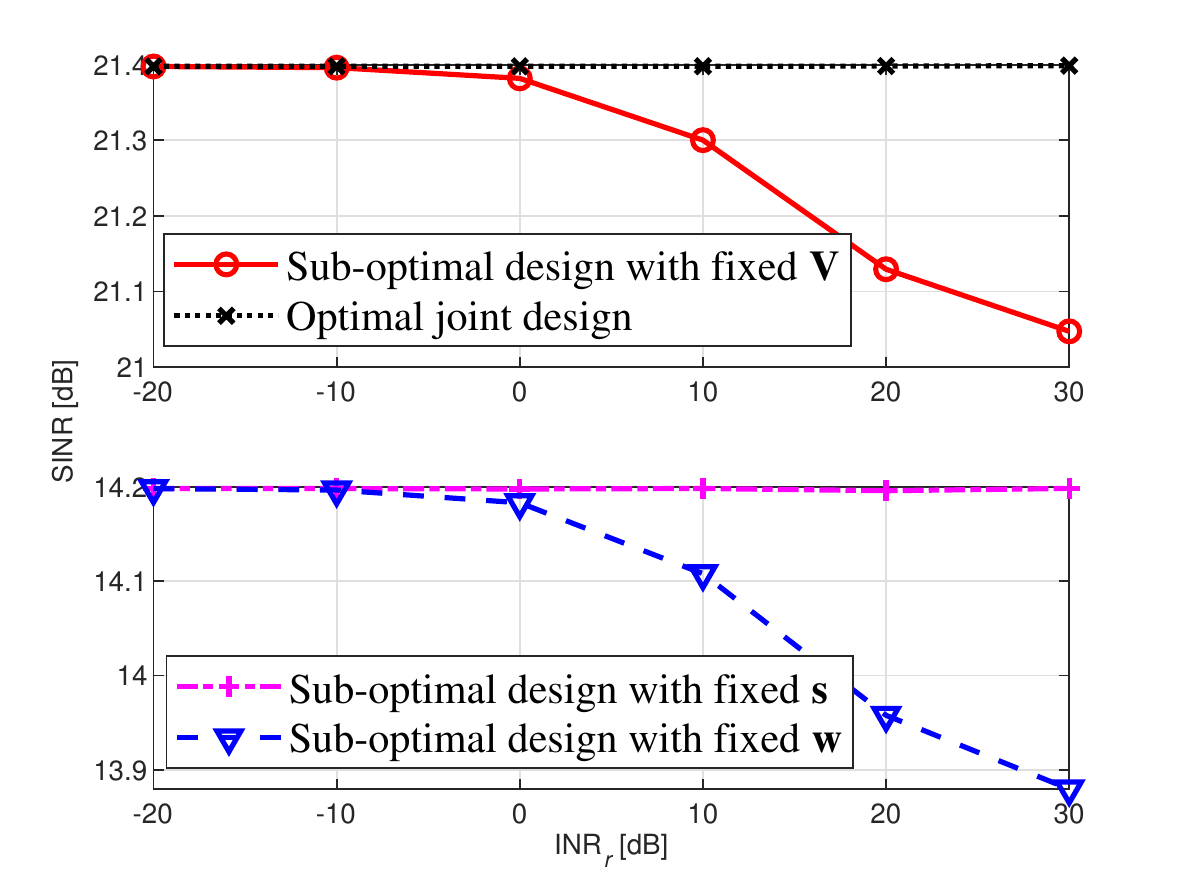}	
	\caption{The radar SINR under different ${\rm INR}_{r}$.}\label{INRr}
\end{figure}

\begin{figure}[!t]
	\centering
	\includegraphics[width=0.43\textwidth]{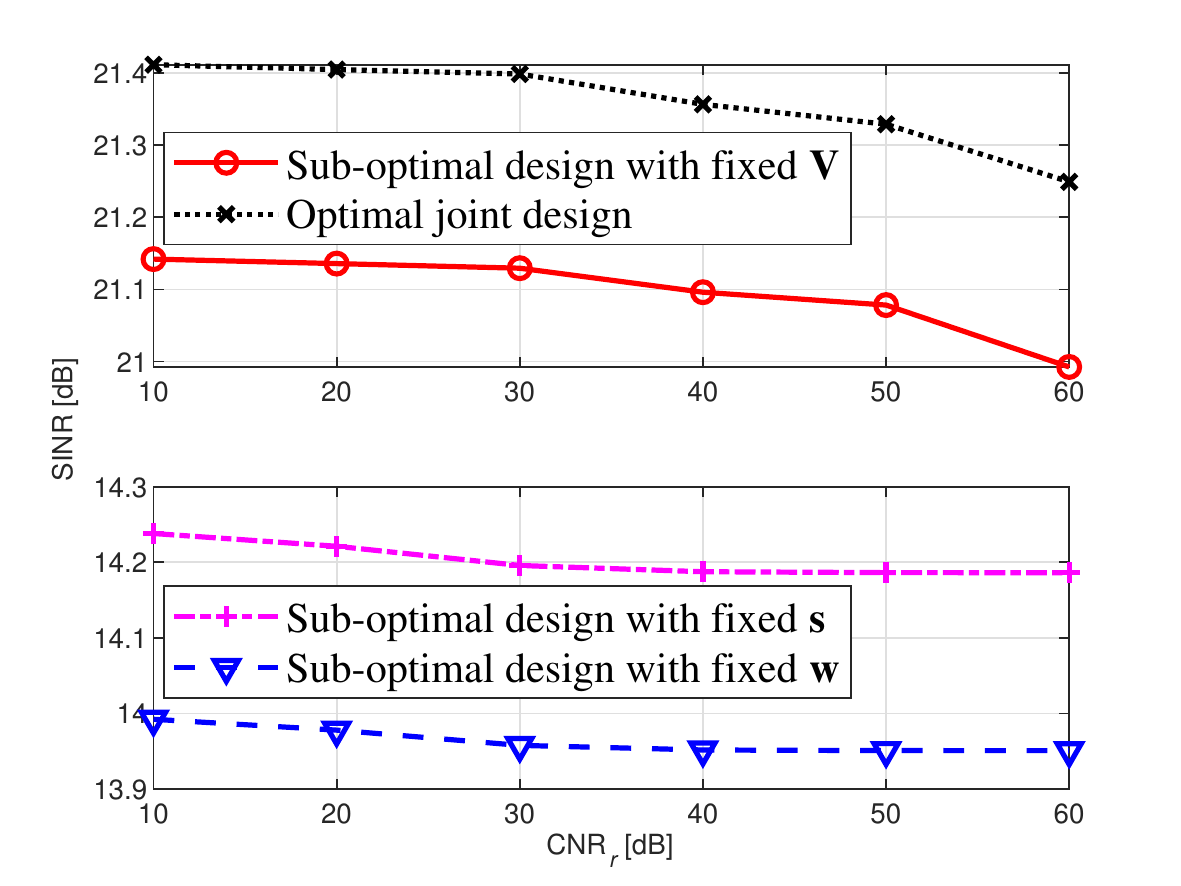}	
	\caption{The radar SINR under different ${\rm CNR}_{r}$.}\label{CNRr}
\end{figure}

Fig. \ref{INRr} and Fig. \ref{CNRr} compare the radar SINR performance of the optimal joint design of $\textbf{V}$, $\textbf{s}$ and $\textbf{w}$ (alternating exploitation of Algorithm 1, Algorithm 2 and formula $(\ref{w_update})$) with the following benchmarks under different ${\rm INR}_{r}$ and ${\rm CNR}_{r}$, respectively.
\begin{itemize}
	\item\emph{Sub-optimal design with fixed \textbf{V}: }We design \textbf{s} and \textbf{w} by iteratively exploiting Algorithm 2 and updating $(\ref{w_update})$. The communication precoder is fixed as ${\textbf{V}_{0}}$.
	\item\emph{Sub-optimal design with fixed \textbf{s}: }We design \textbf{V} and \textbf{w} by iteratively exploiting Algorithm 1 and updating $(\ref{w_update})$. The radar waveform is fixed as $\textbf{s}_0$.
	\item\emph{Sub-optimal design with fixed \textbf{w}: }We design \textbf{s} and \textbf{V} by iteratively exploiting Algorithm 2 and Algorithm 1. The radar space-time filter is fixed as $\textbf{w}_0$.	
\end{itemize}
It is shown that the output SINR of the optimal joint design is the highest and can be maintained in a wide range of ${\rm CNR}_{r}$'s and ${\rm INR}_{r}$'s. We can also observe that the performance of the sub-optimal design with fixed \textbf{V} and the optimal joint design are almost the same when ${\rm INR}_{r}$ is low, which indicates that the communication interference can be ignored in this case. The performance gap between these two designs increases with ${\rm INR}_{r}$. This is because, when ${\rm INR}_{r}$ is high, the optimal joint design can mitigate the communication interference by allocating the transmit power of the BS to the appropriate direction with the precoder design. As expected, the sub-optimal design with fixed \textbf{s} and the sub-optimal design with fixed \textbf{w} perform poorly, mainly due to the lack of designs for \textbf{s} and \textbf{w}, respectively, which is crucial for the radar SINR maximization.

\begin{figure}[!t]
	\centering
	\includegraphics[width=0.43\textwidth]{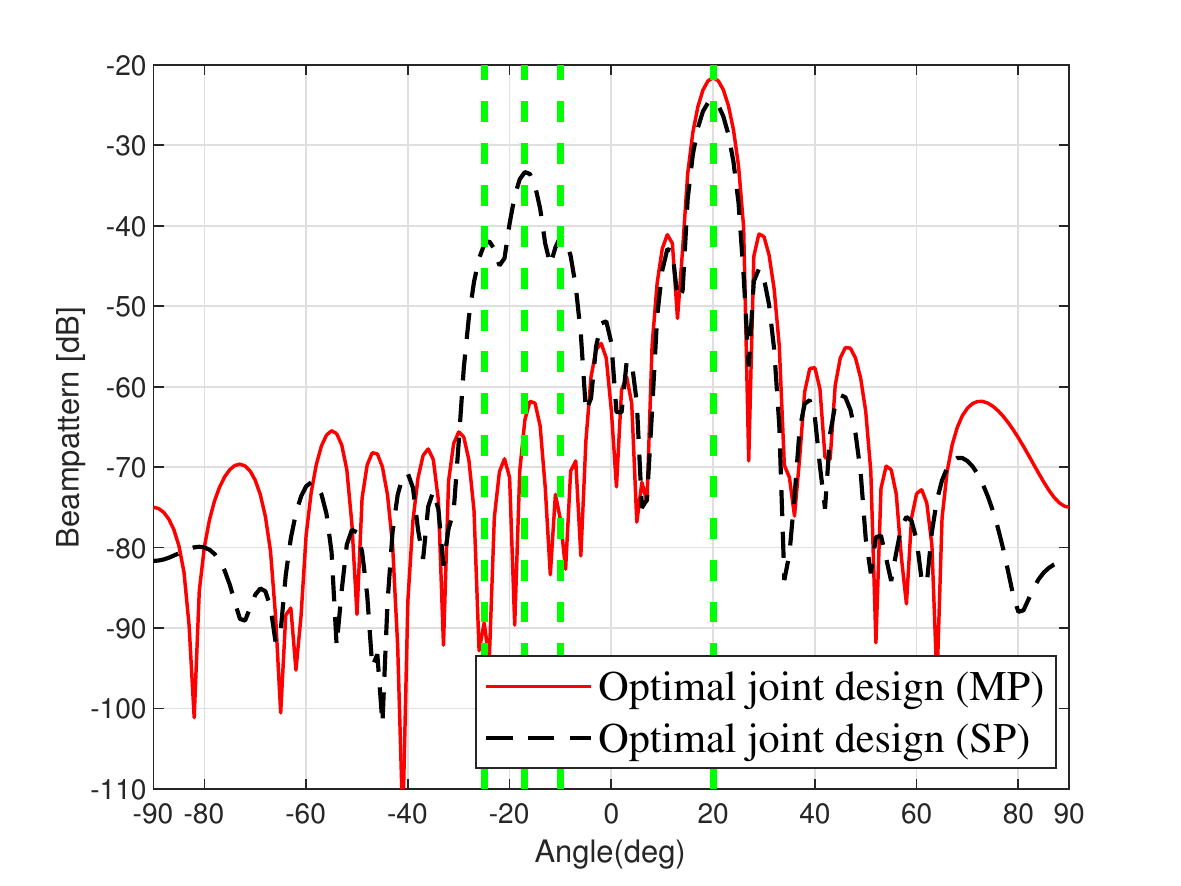}	
	\caption{The radar transceiver beampatterns.}\label{pattern}
\end{figure}

In Fig. \ref{pattern}, we compare the radar beampatterns of the above optimal joint design with MP detection and the optimal joint design with SP detection. The beampattern can be defined as \cite{7169543}
$$P\left( \theta  \right)=\frac{{{\left\| {\textbf{w}^{H}}{\textbf{H}_{0}}\left( \theta  \right)\textbf{s} \right\|}^{2}}}{{{M}_{T}}{{M}_{R}}{{\left\| \textbf{w} \right\|}^{2}}{{\left\| \textbf{s} \right\|}^{2}}}.$$ We can observe that the beampattern for the MP combining peaks at the target and multi-path scatterer location, which illustrates the spatial diversity of multi-path propagation.

\begin{figure}[!t]
	\centering
	\includegraphics[width=0.43\textwidth]{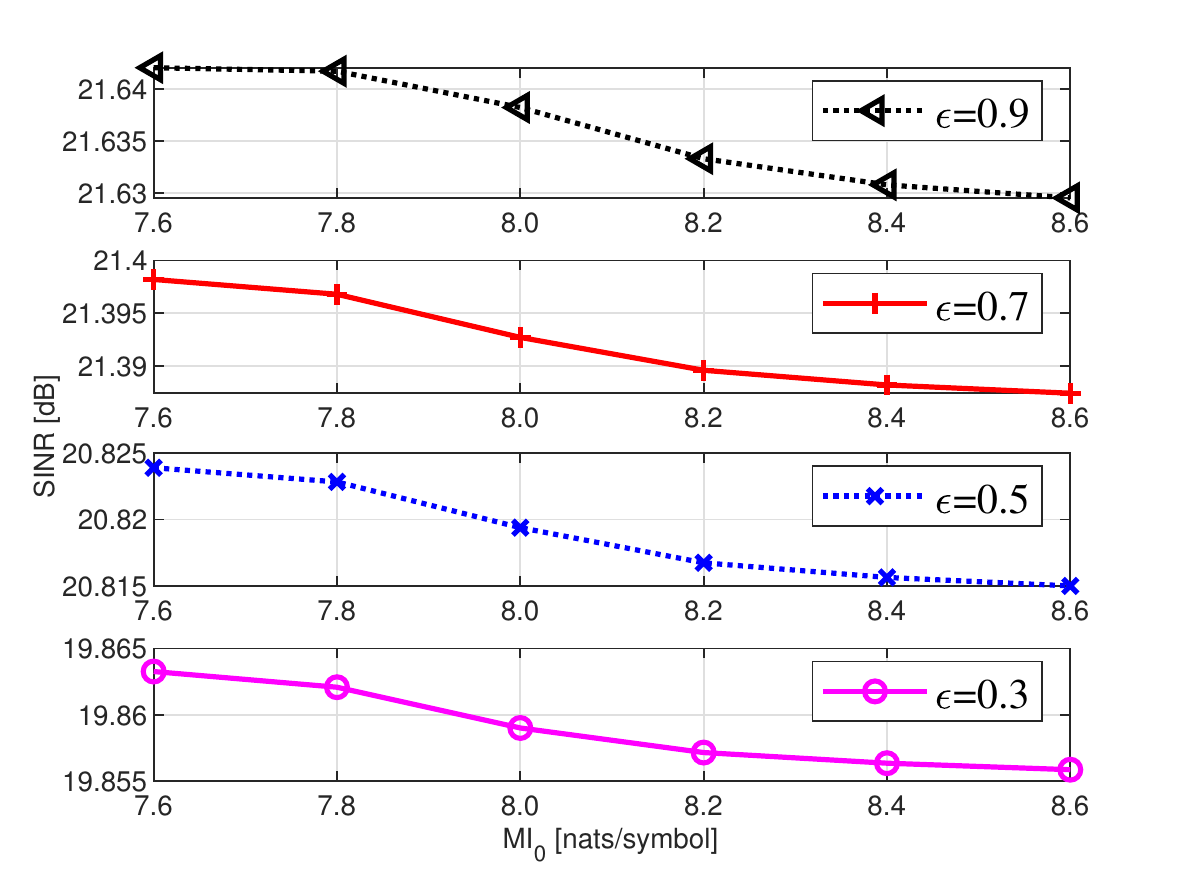}	
	\caption{The radar SINR under different $\epsilon$ and ${\rm MI}_0$.}\label{epsilon}
\end{figure}

Fig. \ref{epsilon} analyzes the radar SINR performance of above optimal joint design under different similarity level $\epsilon$ and ${\rm MI}_0$. $\textbf{s}_0$ is selected as the reference waveform. We can observe that the SINR increases with $\epsilon$ and decreases with ${\rm MI}_0$. This is mainly due to a larger feasible set brought by a larger $\epsilon$ or a smaller ${\rm MI}_0$.

\begin{figure}[!t]	
	\centering
	\includegraphics[width=0.43\textwidth]{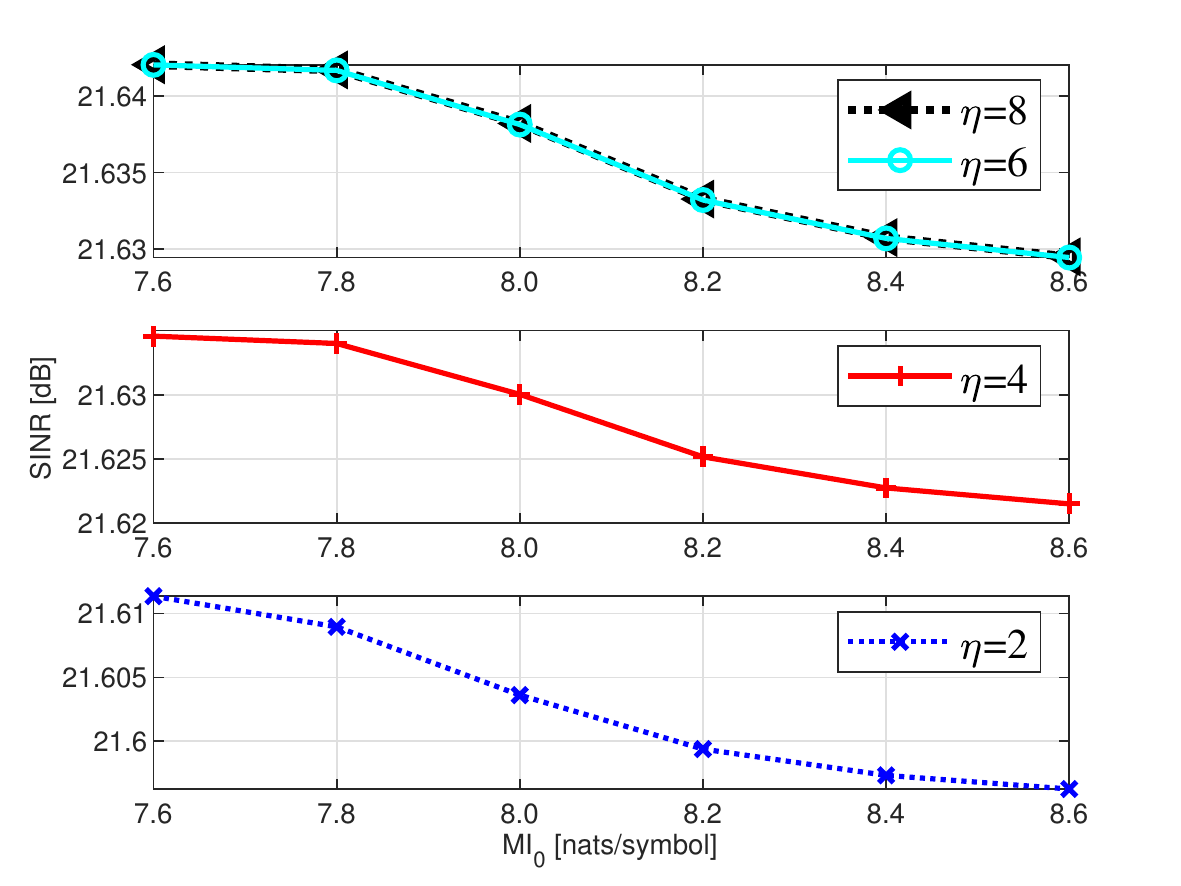}	
	\caption{The radar SINR under different $\eta$ and ${\rm MI}_0$.}\label{papr}
\end{figure}

Finally, in Fig \ref{papr}, we investigate the effect of PAPR constraints on the radar SINR performance. The optimal joint design of $\textbf{V}$, $\textbf{s}$ and $\textbf{w}$ is realized by exploiting Algorithm 1, Algorithm 3 and formula $(\ref{w_update})$ iteratively. As expected, relaxing the PAPR constraints will lead to a larger SINR value. When $\eta$ is larger than 6.0, further increasing it will not significantly improve the radar SINR.

\section{Conclusion}
In this work, we have considered the coexistence of the co-located MIMO radar and single-user MIMO communication systems in multi-path environments. The communication precoder, radar transmit waveform and receive filter have been jointly designed to suppress the mutual interference between the two systems while combining the multi-path signals received by each system. The system design has been conducted with the goal of maximizing the radar SINR, while accounting for the constraints on the transmit power of both systems, radar waveform and the transmission rate at the BS. The formulated non-convex problem has been sub-optimally solved by an iterative algorithm based on the alternating maximization. Simulation results have demonstrated that the proposed algorithm can effectively utilize multi-path signals to achieve better performance and mitigate the interference between the two systems.

\begin{appendices}
\section{Proof of Lemma 1}
Define ${{\tilde{\textbf{a}}}_{q}}\left( \vartheta _{r}^{l},\vartheta _{t}^{l} \right)=\frac{1}{\sqrt{{{N}_{R}}}}{{e}^{-j\pi \left( q-1 \right)\sin \vartheta _{r}^{l}}}{\textbf{a}_{t}}\left( \vartheta _{t}^{l} \right)$, $\boldsymbol{\Delta}_{{{q}_{1}}{{q}_{2}}}^{l}=\tilde{\textbf{a}}_{{{q}_{2}}}^{*}\left( \vartheta _{r}^{l},\vartheta _{t}^{l} \right)\tilde{\textbf{a}}_{{{q}_{1}}}^{T}\left( \vartheta _{r}^{l},\vartheta _{t}^{l} \right)$ and
\begin{equation}
\boldsymbol{\Delta} =\sum\limits_{l=1}^{L}{\sigma _{\upsilon ,l}^{2}\left[ \begin{matrix}
	{\textbf{I}_{D}}\otimes {{\left( \boldsymbol{\Delta} _{11}^{l} \right)}^{T}} & \cdots  & {\textbf{I}_{D}}\otimes {{\left( \boldsymbol{\Delta} _{1{{N}_{R}}}^{l} \right)}^{T}}  \\
	\vdots  & \ddots  & \vdots   \\
	{\textbf{I}_{D}}\otimes {{\left( \boldsymbol{\Delta} _{{{N}_{R}}1}^{l} \right)}^{T}} & \cdots  & {\textbf{I}_{D}}\otimes {{\left( \boldsymbol{\Delta} _{{{N}_{R}}{{N}_{R}}}^{l} \right)}^{T}}  \\
	\end{matrix} \right]}.\label{delta}
\end{equation}
It is easy to verify that the matrix $\boldsymbol{\Delta}$ is Hermitian positive semidefinite. Then, 
the component $\sum\nolimits_{l=1}^{L}{\sigma _{\upsilon ,l}^{2}\textbf{G}_{l}\textbf{V}{\textbf{V}^{H}}{\textbf{G}_{l}^{H}}}$ in ${\rm MI}\left( \textbf{s},\textbf{V} \right)$
can be reformulated as (\ref{H_c^l}), as shown at the top of next page. Subsequently, the function ${{\rm MI}_{k}}\left( \textbf{s},\textbf{V} \right)$ defined in (\ref{MI_k}) can be rewritten as
\newcounter{mytempeqncnt2}
\begin{figure*}[ht]
\normalsize
\setcounter{mytempeqncnt2}{\value{equation}}
\setcounter{equation}{53}
\begin{equation}
\begin{aligned}
\sum\limits_{l=1}^{L}{\sigma _{\upsilon ,l}^{2}\textbf{G}_{l}\textbf{V}{\textbf{V}^{H}}{{ \textbf{G}_{l}^{H} }}}
&=
\sum\limits_{l=1}^{L}{\sigma _{\upsilon ,l}^{2}{\textbf{a}_{r}}\left( \vartheta _{r}^{l} \right)\textbf{a}_{t}^{T}\left( \vartheta _{t}^{l} \right)\textbf{V}{\textbf{V}^{H}}{{\left( {\textbf{a}_{r}}\left( \vartheta _{r}^{l} \right)\textbf{a}_{t}^{T}\left( \vartheta _{t}^{l} \right) \right)}^{H}}}
=\sum\limits_{l=1}^{L}{\sigma _{\upsilon ,l}^{2}
\left[ \begin{matrix}
{\rm tr}\left( \textbf{V}{\textbf{V}^{H}}\boldsymbol{\Delta} _{11}^{l} \right) & \cdots  & {\rm tr}\left( \textbf{V}{\textbf{V}^{H}}\boldsymbol{\Delta} _{1{{N}_{R}}}^{l} \right)  \\
\vdots  & \ddots  & \vdots   \\
{\rm tr}\left( \textbf{V}{\textbf{V}^{H}}\boldsymbol{\Delta} _{{{N}_{R}}1}^{l} \right) & \cdots  & {\rm tr}\left( \textbf{V}{\textbf{V}^{H}}\boldsymbol{\Delta} _{{{N}_{R}}{{N}_{R}}}^{l} \right)  \\
\end{matrix} \right] }\\ 
&=\sum\limits_{l=1}^{L}{\sigma _{\upsilon ,l}^{2}\left[ \begin{matrix}
{\textbf{v}^{T}}\left( {\textbf{I}_{D}}\otimes {{\left( \boldsymbol{\boldsymbol{\Delta}} _{11}^{l} \right)}^{T}} \right){\textbf{v}^{*}} & \cdots  & {\textbf{v}^{T}}\left( {\textbf{I}_{D}}\otimes {{\left( \boldsymbol{\Delta} _{1{{N}_{R}}}^{l} \right)}^{T}} \right){\textbf{v}^{*}}  \\
\vdots  & \ddots  & \vdots   \\
{\textbf{v}^{T}}\left( {\textbf{I}_{D}}\otimes {{\left( \boldsymbol{\Delta} _{{{N}_{R}}1}^{l} \right)}^{T}} \right){\textbf{v}^{*}} & \cdots  & {\textbf{v}^{T}}\left( {\textbf{I}_{D}}\otimes {{\left( \boldsymbol{\Delta} _{{{N}_{R}}{{N}_{R}}}^{l} \right)}^{T}} \right){\textbf{v}^{*}}  \\
\end{matrix} \right]} \\ 
&=\sum\limits_{l=1}^{L}{\sigma _{\upsilon ,l}^{2}\left( {\textbf{I}_{{{N}_{R}}}}\otimes {\textbf{v}^{T}} \right)\left[ \begin{matrix}
{\textbf{I}_{D}}\otimes {{\left( \boldsymbol{\Delta} _{11}^{l} \right)}^{T}} & \cdots  & {\textbf{I}_{D}}\otimes {{\left( \boldsymbol{\Delta} _{1{{N}_{R}}}^{l} \right)}^{T}}  \\
\vdots  & \ddots  & \vdots   \\
{\textbf{I}_{D}}\otimes {{\left( \boldsymbol{\Delta} _{{{N}_{R}}1}^{l} \right)}^{T}} & \cdots  & {\textbf{I}_{D}}\otimes {{\left( \boldsymbol{\Delta} _{{{N}_{R}}{{N}_{R}}}^{l} \right)}^{T}}  \\
\end{matrix} \right]\left( {\textbf{I}_{{{N}_{R}}}}\otimes {\textbf{v}^{*}} \right)}\\
&=\left( {\textbf{I}_{{{N}_{R}}}}\otimes {\textbf{v}^{T}} \right)\boldsymbol{\Delta} \left( {\textbf{I}_{{{N}_{R}}}}\otimes {\textbf{v}^{*}}\right)\label{H_c^l}
\end{aligned}
\end{equation}
\setcounter{equation}{\value{mytempeqncnt2}}
\hrulefill
\end{figure*}
\setcounter{equation}{54}
\begin{equation}
\begin{aligned}
& {{\rm MI}_{k}}\left( \textbf{v} \right)=\log \left| \left( {\textbf{I}_{{{N}_{R}}}}\otimes {\textbf{v}^{T}} \right)\boldsymbol{\Delta} \left( {\textbf{I}_{{{N}_{R}}}}\otimes {\textbf{v}^{*}}\right){{\left( \textbf{R}_{c}^{k}\left( \textbf{s} \right) \right)}^{-1}}+{\textbf{I}_{{{N}_{R}}}} \right| \\ 
& \overset{(\rm a)}{=}\log \left| {{\boldsymbol{\Delta} }^{\frac{1}{2}}}\left( {\textbf{I}_{{{N}_{R}}}}\otimes {\textbf{v}^{*}} \right){{\left( \textbf{R}_{c}^{k}\left( \textbf{s} \right) \right)}^{-1}}\left( {\textbf{I}_{{{N}_{R}}}}\otimes {\textbf{v}^{T}} \right){{\boldsymbol{\Delta} }^{\frac{1}{2}}}+{\textbf{I}_{{{N}_{R}}{{N}_{T}}D}} \right|\\
&\overset{(\rm b)}{=}\log \left| \textbf{C}{{\left( {\textbf{E}_{k}}\left( \textbf{v} \right) \right)}^{-1}}{\textbf{C}^{H}} \right|, 
\end{aligned}
\end{equation}
where the procedure (a) is due to $\left|  \textbf{I}+\textbf{A}_1\textbf{A}_2 \right|=\left| \textbf{I}+\textbf{A}_2\textbf{A}_1\right|$, the procedure (b) uses the inversion lemma of a partitioned matrix, and the matrix ${\textbf{E}_{k}}\left( \textbf{v} \right)$ is given by (\ref{E_k}) \cite{horn2012matrix}. Thus, the function ${\rm MI}\left( \textbf{v} \right)=\frac{1}{{\tilde{K}}}\sum\nolimits_{k=1}^{\tilde{K}} {{\rm MI}_{k}}\left( \textbf{v} \right)$ is equivalent to ${\rm MI}\left( \textbf{s},\textbf{V} \right)$.

\section{Proof of Lemma 2}
Since the function $\log \left| \textbf{C}{{\left( {\textbf{E}_{k}}\left( \textbf{v} \right) \right)}^{-1}}{\textbf{C}^{H}} \right|$ in Lemma 1 is convex w.r.t. ${\textbf{E}_{k}}\left( \textbf{v} \right)$ \cite{7895135}, we have the following first-order condition
\begin{equation}
\begin{aligned}
\log \left| \textbf{C}{{\left( {\textbf{E}_{k}}\left( \textbf{v} \right) \right)}^{-1}}{\textbf{C}^{H}} \right|\ge &\log \left|\textbf{C}{{\left( {\textbf{E}_{k}}\left( {\bar{\textbf{v}}} \right) \right)}^{-1}}{\textbf{C}^{H}} \right|\\
&+{\rm tr}\left( {{\boldsymbol{\Gamma} }_{k}}\left( {\bar{\textbf{v}}} \right)\left( {\textbf{E}_{k}}\left( \textbf{v} \right)-{\textbf{E}_{k}}\left( {\bar{\textbf{v}}} \right) \right) \right),
\end{aligned}
\end{equation}
where
\begin{equation}
{{\boldsymbol{\Gamma} }_{k}}\left( {\bar{\textbf{v}}} \right)=-{{\left( {\textbf{E}_{k}}\left( {\bar{\textbf{v}}} \right) \right)}^{-1}}{\textbf{C}^{H}}{{\left( \textbf{C}{{\left( {\textbf{E}_{k}}\left( {\bar{\textbf{v}}} \right) \right)}^{-1}}{\textbf{C}^{H}} \right)}^{-1}}\textbf{C}\big( {\textbf{E}_{k}}\left( {\bar{\textbf{v}}} \right) \big)^{-1}
\end{equation}
is the gradient of $\log \left| \textbf{C}{{\left( {\textbf{E}_{k}}\left( \textbf{v} \right) \right)}^{-1}}{\textbf{C}^{H}} \right|$ at ${\textbf{E}_{k}}\left( {\bar{\textbf{v}}} \right)$. Thus, the constraint $(\rm \ref{com_MI})$ can be converted to
\begin{equation}
\begin{aligned}
&\frac{1}{{\tilde{K}}}\sum\limits_{k=1}^{{\tilde{K}}}{\Big(\log \left|\textbf{C}{{\left( {\textbf{E}_{k}}\left( {\bar{\textbf{v}}} \right) \right)}^{-1}}{\textbf{C}^{H}} \right|+{\rm tr}\left( {{\boldsymbol{\Gamma} }_{k}}\left( {\bar{\textbf{v}}} \right)\left( {\textbf{E}_{k}}\left( \textbf{v} \right)-{\textbf{E}_{k}}\left( {\bar{\textbf{v}}} \right) \right) \right)\Big)}\\
&\ge {\rm MI}_0.
\end{aligned}\label{com_MI_appro}
\end{equation}

Let ${{\boldsymbol{\Gamma} }_{k}}\left( {\bar{\textbf{v}}} \right)$ be partitioned as $${{\boldsymbol{\Gamma}}_{k}}\left( {\bar{\textbf{v}}} \right)=\left[ \begin{matrix}
{{\boldsymbol{\Gamma}}_{k}^{11}}\left( {\bar{\textbf{v}}} \right) & {{\boldsymbol{\Gamma} }_{k}^{12}}\left( {\bar{\textbf{v}}} \right)  \\
{{\boldsymbol{\Gamma} }_{k}^{21}}\left( {\bar{\textbf{v}}} \right) & {{\boldsymbol{\Gamma}}_{k}^{22}}\left( {\bar{\textbf{v}}} \right)  \\
\end{matrix} \right].$$ Then, we can obtain that
$$
\begin{aligned}
{\rm tr}\left( {{\boldsymbol{\Gamma}}_{k}}\left( {\bar{\textbf{v}}} \right){\textbf{E}_{k}}\left( \textbf{v} \right) \right)&={\rm tr}\left( {{\boldsymbol{\Gamma}}_{k}^{11}}\left( {\bar{\textbf{v}}} \right) \right)+{\rm tr}\left( {{\boldsymbol{\Gamma}}_{k}^{22}}\left( {\bar{\textbf{v}}} \right)\textbf{R}_{c}^{k}\left( \textbf{s} \right) \right)\\
&+2\mathcal{R}\left( {\rm tr}\left( {{\boldsymbol{\Gamma}}_{k}^{12}}\left( {\bar{\textbf{v}}} \right)\left( {\textbf{I}_{{{N}_{R}}}}\otimes {\textbf{v}^{T}} \right){{\boldsymbol{\Delta} }^{\frac{1}{2}}} \right) \right)\\
&+{\rm tr}\left( {{\boldsymbol{\Gamma}}_{k}^{22}}\left( {\bar{\textbf{v}}} \right)\left( {\textbf{I}_{{{N}_{R}}}}\otimes {\textbf{v}^{T}} \right)\boldsymbol{\Delta} \left( {\textbf{I}_{{{N}_{R}}}}\otimes {\textbf{v}^{*}} \right) \right).
\end{aligned}$$
Thus, the constraint $(\ref{com_MI_appro})$ can be rewritten as
\begin{equation}
\begin{aligned}
&2\mathcal{R}\left( {\rm tr}\left( {{\boldsymbol{\Gamma}}_{12}}\left( {\bar{\textbf{v}}} \right)\left( {\textbf{I}_{{{N}_{R}}}}\otimes {\textbf{v}^{T}} \right){{\boldsymbol{\Delta} }^{\frac{1}{2}}} \right) \right)\\&+{\rm tr}\left( {{\boldsymbol{\Gamma}}_{22}}\left( {\bar{\textbf{v}}} \right)\left( {\textbf{I}_{{{N}_{R}}}}\otimes {\textbf{v}^{T}} \right)\boldsymbol{\Delta} \left( {\textbf{I}_{{{N}_{R}}}}\otimes {\textbf{v}^{*}} \right) \right)
\ge \overline{\rm MI}\left( {\bar{\textbf{v}}} \right)\label{com_MI_app}
\end{aligned}
\end{equation}
where 
${{\boldsymbol{\Gamma}}_{12}}\left( {\bar{\textbf{v}}} \right)=\frac{1}{{\tilde{K}}}\sum\nolimits_{k=1}^{{\tilde{K}}}{{{\boldsymbol{\Gamma}}_{k}^{12}}\left( {\bar{\textbf{v}}} \right)}$, ${{\boldsymbol{\Gamma}}_{22}}\left( {\bar{\textbf{v}}} \right)=\frac{1}{{\tilde{K}}}\sum\nolimits_{k=1}^{{\tilde{K}}}{{{\boldsymbol{\Gamma}}_{k}^{22}}\left( {\bar{\textbf{v}}} \right)}$ and
\begin{equation}
\begin{aligned}
&\overline{{\rm MI}}\left( {\bar{\textbf{v}}} \right)={{\rm MI}_{0}}-\frac{1}{{\tilde{K}}}\sum\limits_{k=1}^{{\tilde{K}}}\Big(\log \left| \textbf{C}{{\left( {\textbf{E}_{k}}\left( \bar{\textbf{v}} \right) \right)}^{-1}}{\textbf{C}^{H}} \right|-\\
&{\rm tr}\left( {{\boldsymbol{\Gamma}}_{k}}\left( {\bar{\textbf{v}}} \right){\textbf{E}_{k}}\left( {\bar{\textbf{v}}} \right) \right)+{\rm tr}\left( {{\boldsymbol{\Gamma}}_{k}^{11}}\left( \bar{\textbf{v}} \right) \right)+{\rm tr}\left( {{\boldsymbol{\Gamma}}_{k}^{22}}\left( \bar{\textbf{v}} \right)\textbf{R}_{c}^{k}\left( \textbf{s} \right) \right) \Big).
\end{aligned}\label{MI_bar}
\end{equation}

By using ${\rm tr}\big( {\textbf{A}_1^{T}}\textbf{A}_2 \big)={{\rm vec}^{T}}\left( \textbf{A}_1 \right){\rm vec}\left( \textbf{A}_2 \right)$ and ${\rm tr}\left( \textbf{A}_1\textbf{A}_2\textbf{A}_3\textbf{A}_4 \right)={{\rm vec}^{T}}\left( \textbf{A}_4 \right)\left( \textbf{A}_1\otimes {\textbf{A}_3^{T}} \right){\rm vec}\big( {\textbf{A}_2^{T}} \big)$, we can convert (\ref{com_MI_app}) to
\begin{equation}
\begin{aligned}
&{\textbf{v}^{H}}{\textbf{P}^{H}}\left( {{\boldsymbol{\Gamma} }_{22}}\left( {\bar{\textbf{v}}} \right)\otimes {{\boldsymbol{\Delta} }^{T}} \right)\textbf{Pv}+2\mathcal{R}\left( {{\rm vec}^{T}}\left( {{\boldsymbol{\Delta} }^{\frac{1}{2}}}{{\boldsymbol{\Gamma}}_{12}}\left( {\bar{\textbf{v}}} \right) \right)\textbf{Pv} \right)\\
&\ge \overline{\rm MI}\left( {\bar{\textbf{v}}} \right),
\end{aligned}
\end{equation}
where $\textbf{P}={{\left[ {\textbf{P}_{1}},{\textbf{P}_{2}},\ldots ,{\textbf{P}_{{{N}_{R}}}} \right]}^{T}}$ with ${\textbf{P}_{i}}=\tilde{\textbf{e}}_{i}^{T}\otimes {\textbf{I}_{{{N}_{T}}D}}$ denoting a ${{N}_{T}}D\times {{N}_{T}}D{{N}_{R}}$ matrix for $i=1,2,\ldots ,{{N}_{R}}$ and $\tilde{\textbf{e}}\left( n \right)\in {{\mathbb{C}}^{{N}_R}}$ being a direction vector similar to ${\textbf{e}}\left( n \right)$. As a result, we can successively approximate the original constraint $(\rm \ref{com_MI})$ by
\begin{equation}
{\textbf{v}^{H}}{{\bar{\boldsymbol{\Gamma}}}_{22}}\left( {\bar{\textbf{v}}} \right)\textbf{v}-2\mathcal{R}\left( {{{\bar{\boldsymbol{\Gamma}}}}_{12}}\left( {\bar{\textbf{v}}} \right)\textbf{v} \right)\le -\overline{\rm MI}\left( {\bar{\textbf{v}}} \right),
\end{equation}
where
\begin{equation} 
{{\bar{\boldsymbol{\Gamma}}}_{22}}\left( {\bar{\textbf{v}}} \right)=-{\textbf{P}^{H}}\left( {{\boldsymbol{\Gamma}}_{22}}\left( {\bar{\textbf{v}}} \right)\otimes {\boldsymbol{\boldsymbol{\Delta} }^{T}} \right)\textbf{P}\label{tao_22}
\end{equation}
and
\begin{equation} 
{{\bar{\boldsymbol{\Gamma} }}_{12}}\left( {\bar{\textbf{v}}} \right)={{\rm vec}^{T}}\left( {{\boldsymbol{\Delta} }^{\frac{1}{2}}}{{\boldsymbol{\Gamma} }_{12}}\left( {\bar{\textbf{v}}} \right) \right)\textbf{P}. \label{tao_12}
\end{equation}

Since ${\textbf{E}_{k}}\left( \textbf{v} \right)$ is positive definite, ${{\boldsymbol{\Gamma}}_{k}}\left( {\bar{\textbf{v}}} \right)$, ${{\boldsymbol{\Gamma}}_{k}^{22}}\left( {\bar{\textbf{v}}} \right)$ and ${{\boldsymbol{\Gamma}}_{22}}\left( {\bar{\textbf{v}}} \right)$ are all negative semidefinite. Hence, ${{\bar{\boldsymbol{\Gamma}}}_{22}}\left( {\bar{\textbf{v}}} \right)$ is positive semidefinite and the constraint is convex. 

\section{Proof of Proposition 4}
In the case of LoS detection, we have ${\rm rank}(\tilde{\boldsymbol{\Psi}})=1$.
The Lagrangian of $\mathcal{P}\text{2.2}\big( \bar{\textbf{S}}\big)$ can be given by
\begin{equation}\nonumber
\begin{aligned}
&\mathcal{L}\big( \tilde{\textbf{S}}_2,{\left\{ {{\lambda }_{k}} \right\}}_{k=1}^{3},\tilde{\textbf{Y}} \big)={\rm tr}\big( \tilde{\textbf{S}}_2 \big)+{{\lambda }_{1}}\big( \tilde{p}{\rm tr}\big( \tilde{\textbf{R}}\tilde{\textbf{S}}_2 \big)+\tilde{p}{r(\textbf{V})}-\\
&{\rm tr}\big( \tilde{\boldsymbol{\Psi} }\tilde{\textbf{S}}_2 \big) \big)+{{{\lambda }_{2}}\big( {\rm tr}\big( {{\hat{\boldsymbol{\Gamma} }\left( {\bar{\textbf{S}}} \right)}}\tilde{\textbf{S}}_2 \big)-\widehat{\rm MI}\left( {\bar{\textbf{S}}} \right) \big)}+
{{\lambda }_{3}}\Big({\rm tr}\Big(\Big({\textbf{I}}_{KM_T}\\
&
 -\frac{{\textbf{s}_{0}}\textbf{s}_{0}^{H}}{{{P}_{R}}}\Big)\tilde{\textbf{S}}_2 \Big)-\epsilon {{P}_{R}}\Big)
-{\rm tr}\big( \tilde{\textbf{Y}}\tilde{\textbf{S}}_2 \big) 
={\rm tr}\Big( \Big( {\textbf{I}_{KM_T}}+{{\lambda }_{1}}\tilde{p}\tilde{\textbf{R}}-\\
&{{\lambda }_{1}}\tilde{\boldsymbol{\Psi} }+{{\lambda }_{2}}{{\hat{\boldsymbol{\Gamma} }\left( {\bar{\textbf{S}}} \right)}}+{{\lambda }_{3}}\Big({\textbf{I}}_{KM_T}-\frac{{\textbf{s}_{0}}\textbf{s}_{0}^{H}}{{{P}_{R}}}\Big)-\tilde{\textbf{Y}} \Big)\tilde{\textbf{S}}_2 \Big)+{{\lambda }_{1}}\tilde{p}{r(\textbf{V})}\\
&-{{\lambda }_{2}}\widehat{\rm MI}\left( {\bar{\textbf{S}}} \right)-{{\lambda }_{3}}\epsilon {{P}_{R}},
\end{aligned}
\end{equation}
where ${{\lambda }_{1}}$, ${{\lambda }_{2}} $, ${{\lambda }_{3}} $ and $\tilde{\textbf{Y}}$ denote the Lagrange multipliers. Setting the gradient of $\mathcal{L}\left( \tilde{\textbf{S}}_2,{\left\{ {{\lambda }_{k}} \right\}}_{k=1}^{3},\tilde{\textbf{Y}} \right)$ to zero, we obtain
\begin{equation}\nonumber
{\textbf{I}_{KM_T}}+{{\lambda }_{1}}\tilde{p}\tilde{\textbf{R}}-{{\lambda }_{1}}\tilde{\boldsymbol{\Psi} }+{{\lambda }_{2}}{{\hat{\boldsymbol{\Gamma} }\left( {\bar{\textbf{S}}} \right)}}+{{\lambda }_{3}}\Big({\textbf{I}}_{KM_T}-\frac{{\textbf{s}_{0}}\textbf{s}_{0}^{H}}{{{P}_{R}}}\Big)-\tilde{\textbf{Y}}=\textbf{0}.
\end{equation}
Define $\boldsymbol{\Xi} ={\textbf{I}_{KM_T}}+{{\lambda }_{1}}\tilde{p}\tilde{\textbf{R}}+{{\lambda }_{2}}{{\hat{\boldsymbol{\Gamma} }\left( {\bar{\textbf{S}}} \right)}}+{{\lambda }_{3}}\Big({\textbf{I}}_{KM_T}-\frac{{\textbf{s}_{0}}\textbf{s}_{0}^{H}}{{{P}_{R}}}\Big)=\tilde{\textbf{Y}}+{{\lambda}_{1}}\tilde{\boldsymbol{\Psi}}$. We have $\boldsymbol{\Xi} \succ \textbf{0}$ and ${\rm rank}\left( \boldsymbol{\Xi}  \right)=K{{M}_{T}}$. On the one hand, since $\tilde{\textbf{S}}_2\ne \textbf{0}$ and $\tilde{\textbf{Y}}\tilde{\textbf{S}}_2=\textbf{0}$, we can easily infer that ${{\lambda }_{1}}>0$ and ${\rm rank}\left( {\tilde{\textbf{Y}}} \right)\le K{{M}_{T}}-1$. On the other hand, since ${\rm rank}\left( \boldsymbol{\Xi}  \right)={\rm rank}\big( {{\lambda}_{1}}\tilde{\boldsymbol{\Psi} }+\tilde{\textbf{Y}} \big)\le {\rm rank}\big( {{\lambda }_{1}}\tilde{\boldsymbol{\Psi} } \big)+{\rm rank}\left( {\tilde{\textbf{Y}}} \right)=1+{\rm rank}\left( {\tilde{\textbf{Y}}} \right)$, we have ${\rm rank}\left( {\tilde{\textbf{Y}}} \right)\ge K{{M}_{T}}-1$. Thus, we obtain ${\rm rank}\left( {\tilde{\textbf{Y}}} \right)= K{{M}_{T}}-1$. Combined with $\tilde{\textbf{Y}}\tilde{\textbf{S}}_2=\textbf{0}$, it follows that ${\rm rank}\big( {\tilde{\textbf{S}}}_2 \big)\le 1$. Based on $\tilde{\textbf{S}}_2\ne \textbf{0}$, it can be inferred that ${\rm rank}\big( {\tilde{\textbf{S}}}_2 \big)=1$.

\end{appendices}

\section*{}
\bibliography{mycite}
\end{document}